\documentclass[journal=jctcce, manuscript=article]{achemso}

\usepackage[version=3]{mhchem} % Formula subscripts using \ce{}
\usepackage{mathtools}
\usepackage{amsmath} % math formulas
\usepackage{amssymb} % math formulas
\usepackage{amsfonts} % math formulas
\usepackage{dsfont} % math formulas
\usepackage{graphicx} % for figures
\graphicspath{ {./images/} } %images path
\usepackage{xcolor} % for code type writting
\usepackage{stmaryrd} % for math formulas
\usepackage{enumitem} % for enumera
\usepackage[ruled,vlined]{algorithm2e} % for algorithm pseudo-code
\usepackage{xr} % To reference the various sections and objects of the SI tex document.
\usepackage{todonotes} % for writting notes

\definecolor{light-gray}{gray}{0.95}

\author{Thomas Pigeon}
\affiliation[Matherials]
{MATHERIALS team-project, Inria Paris, 2 Rue Simone Iff, 75012 Paris, France}
\email{thomas.pigeon@inria.fr}
\alsoaffiliation[Cermics]
{CERMICS, École des Ponts ParisTech, 6-8 Avenue Blaise Pascal, 77455,Marne-la-Vallée, France}
\alsoaffiliation[IFPEN Solaize]
{IFP Energies Nouvelles, Rond-Point de l’Echangeur de Solaize, BP 3, 69360 Solaize, France}

\author{Gabriel Stoltz}
\affiliation[Cermics]
{CERMICS, École des Ponts ParisTech, 6-8 Avenue Blaise Pascal, 77455,Marne-la-Vallée, France}
\alsoaffiliation[Matherials]
{MATHERIALS team-project, Inria Paris, 2 Rue Simone Iff, 75012 Paris, France}

\author{Manuel Corral-Valero}
\affiliation[IFPEN Solaize]
{IFP Energies Nouvelles, Rond-Point de l’Echangeur de Solaize, BP 3, 69360 Solaize, France}

\author{Ani Anciaux-Sedrakian}
\affiliation[IFPEN Rueil]
{IFP Energies Nouvelles, 1 et 4 avenue de Bois-Préau, F-92852 Rueil-Malmaison Cedex, France}

\author{Maxime Moreaud}
\affiliation[IFPEN Solaize]
{IFP Energies Nouvelles, Rond-Point de l’Echangeur de Solaize, BP 3, 69360 Solaize, France}

\author{Tony Lelièvre}
\affiliation[Cermics]
{CERMICS, École des Ponts ParisTech, 6-8 Avenue Blaise Pascal, 77455,Marne-la-Vallée, France}
\alsoaffiliation[Matherials]
{MATHERIALS team-project, Inria Paris, 2 Rue Simone Iff, 75012 Paris, France}
\email{tony.lelievre@enpc.fr}

\author{Pascal Raybaud}
\affiliation[IFPEN Solaize]
{IFP Energies Nouvelles, Rond-Point de l’Echangeur de Solaize, BP 3, 69360 Solaize, France}
\email{raybaud@ifpen.fr}

\title{Computing Surface Reaction Rates by Adaptive Multilevel Splitting Combined with Machine Learning and Ab Initio Molecular Dynamics}

\abbreviations{AMS,DFT,VASP}

\begin{document}

%\begin{tocentry}
%
%Some journals require a graphical entry for the Table of Contents.
%This should be laid out ``print ready'' so that the sizing of the
%Inside the \texttt{tocentry} environment, the font used is Helvetica
%8\,pt, as required by \emph{Journal of the American Chemical
%The surrounding frame is 9\,cm by 3.5\,cm, which is the maximum
%permitted for  \emph{Journal of the American Chemical Society}
%graphical table of content entries. The box will not resize if the
%content is too big: instead it will overflow the edge of the box.
%
%This box and the associated title will always be printed on a
%separate page at the end of the document.
%
%\end{tocentry}
\newpage 
\begin{abstract}
Computing accurate rate constants for catalytic events occurring at the surface of a given material represents a challenging task with multiple potential applications in chemistry.    
To address this question, we propose an approach based on a combination of the rare event sampling method called Adaptive Multilevel Splitting (AMS) and ab initio molecular dynamics (AIMD). The AMS method requires a one dimensional reaction coordinate to index the progress of the transition. Identifying a good reaction coordinate is difficult, especially for high dimensional problems such a those encountered in catalysis. We probe various approaches to build reaction coordinates such as Support Vector Machine and path collective variables. The AMS is implemented so as to communicate with a DFT-plane wave code. A relevant case study in catalysis: the change of conformation and the dissociation of a water molecule chemisorbed on the (100) $\gamma$-alumina surface is used to evaluate our approach. The calculated rate constants and transition mechanisms are discussed and compared to those obtained by a conventional static approach based on the Eyring-Polanyi equation with harmonic approximation. It is revealed that the AMS method may provide rate constants which are smaller than the static approach by up to two orders of magnitude due to entropic effects involved in the chemisorbed water.
\end{abstract}

\newpage 
\section{Introduction}

The determination of chemical reaction rate constants is of tremendous importance to better understand and quantify the kinetics of molecular transformations. This can be a challenging task, especially in catalysis where multiple elementary steps are involved for one targeted reaction. Evaluating each of them by experimental methods being often out of reach, an alternative lies in the theoretical modeling of each of them. Thanks to the significant increase of computational resources, quantum simulation approaches are widely used nowadays to address numerous catalytic systems involved in petrochemistry, fine chemistry and biomass conversion\cite{Broadbelt2000, Chizallet2014, Chen2020a, Piccini2022}.  

However, at the simulation time scale, such chemical transformations are rare events. The typical time step for the integration of stochastic dynamics modeling the evolution of the system is of the order of $10^{-15}$ s, while the frequency of chemical reactions lies in the range from $10^{0}$ to $10^{12} \; \mathrm{s}^{-1}$. Moreover, to accurately simulate catalytic activation of chemical bond breaking and formation, the simulation must include the explicit treatment of valence electrons and the quantum chemical calculation of the Hellmann-Feynman forces for each step of the dynamics\cite{Feynman1939}. Such an \textit{ab initio} molecular dynamics (AIMD) approach becomes so computationally demanding that it is generally impossible to simulate a trajectory that is long enough to observe multiple reaction events, allowing the accurate quantification of rate constants.

Theoretical approaches most commonly used to explore chemical transformations are based on transition state theory (TST)\cite{Eyring1935}. Within this formalism, the reactant and product are considered to be separated in phase space by a dynamical bottleneck \cite{Bennett1977}, which can be characterized as a surface in the configuration space. For a reaction with only one reactive path and only one energy barrier to cross, assuming momenta are not relevant for the transition process, this surface should contain the first order saddle points. The term transition state (TS) is versatile as sometimes it refers to a first order saddle point and sometimes to a isocommittor surface, as defined by IUPAC\cite{TS_goldbook2014}. Considering TS to be surfaces, the reaction rate can be approximated as the frequency at which this surface is crossed. The most common approach to compute the reaction rate constant is called harmonic TST (hTST) as it allows to reduce general TST expression into the "generalized" Eyring--Polanyi equation thanks to harmonic approximation of the potential energy surface\cite{Haenggi1990, Eyring1935, Evans1935, Wigner1938}.
\begin{equation}
\label{Eyring_Polanyi}
k_\mathrm{hTST} = \kappa(T) \frac{k_\mathrm{B}T}{h} \mathrm{e}^{-\frac{\Delta G^{\ddagger}}{k_\mathrm{B}T}},
\end{equation}
where $\Delta G^{\ddagger}$ is the free energy of activation computed as the difference of the free energy of the metastable basin and of the transition state, $k_\mathrm{B}$ the Boltzmann constant, $h$ the Planck constant, $T$ the temperature and $\kappa(T)$ the transmission coefficient. 
This last quantity has to be between 0 and 1 and accounts for the recrossing of the surface, as discussed later on. The free energy of activation is approximated via an harmonic approximation around the saddle point and the minima. Although hTST is one of the most widely used method to determine activation free energies and the rate constants of chemical events, particularly catalytic ones, it suffers from some weaknesses. Among them, the harmonic approximation of the potential energy surfaces as well as the determination of the prefactor $\kappa$ in~\eqref{Eyring_Polanyi} might be questionable. In general when the entropy of the metastable state and the transition state differ by a non negligible amount, the harmonic approximation can lead to significant errors. This can occur in various systems of interest to catalysis such as solid--liquid interfaces, zeolites, porous solids and supported nano-particles\cite{Collinge2020}.
More general expression for the TST rate, using a one dimensional reaction coordinate\cite{Bennett1977, Chandler1978} and relying on sampling methods to estimate free energies\cite{Chipot2007, Rousset2010, Yang2019}, were proposed to overcome some limitations of~\eqref{Eyring_Polanyi}. However, TST reaction rates contain a transmission coefficient $\kappa \in (0,1]$, accounting for the recrossing of the transition state surface, which is rather difficult to evaluate and which explains why bare TST overestimates the transition rate.\cite{Wigner1938, Horiuti1938, Keck1962, Chandler1978, Bennett1977, VandenEijnden2005}.

There are of course alternative approaches to TST. A first one is based on the evolution of a time correlation function\cite{Miller1983, Chandler1978} which found applications in Transition Path Sampling (TPS)\cite{Dellago2003} or other approaches such as the recent work relying on Onsager--Machlup path probability distribution of Ref. \citenum{Mandelli2020}. Another alternative, that we will use in the present work, is provided by approaches based on the Hill relation\cite{Hill2012}:
\begin{equation}
\label{Hill_relation}
k_\mathrm{Hill} = \Phi_R p_{R \rightarrow P}(\partial R),
\end{equation}
where $\Phi_R$ is the flux of trajectories leaving the reactant state $R$ and the committor probability at the boundary $p_{R \rightarrow P}(\partial R)$, the probability of reaching the product state $P$ before returning to $R$ starting from the boundary $\partial R$. In other words, this relation states that the rate constant is the average rate at which the system attempts to leave the initial state times the probability of success. Relation~\eqref{Hill_relation} has been proven correct assuming that the reactant state $R$ is metastable for systems evolving according to the overdamped Langevin dynamics\cite{Baudel2020} or Langevin dynamics\cite{Lelievre2022}. The Hill relation is used in various approaches corresponding to so-called path sampling methods such as Transition Interface Sampling (TIS)\cite{Erp2003}, Forward Flux Sampling (FFS)\cite{Allen2005}, Weighted Ensembles (WE)\cite{Huber1996} and Adaptive Multi-level Splitting (AMS)\cite{Cerou2007}. All these methods are designed to compute the probability $p_{R \rightarrow P}(\partial R)$, which is the most difficult object to evaluate in~\eqref{Hill_relation}. As a side product, these methods sample some reactive trajectories. 

All the methods described previously have different precisions and computational efficiencies. On the one hand, the hTST approach is, by far, the most inexpensive methodology in terms of computational resources but, as mentioned above, it may lead to significant errors. On the other hand, the computational cost of enhanced sampling methods to estimate free energies is not negligible. The Hill relation has the advantage of being exact compared to TST approach but the required computational cost can be high depending on the method to compute the probability $p_{R \rightarrow P}(\partial R)$. Moreover, numerical methods sampling reactive trajectories offer the possibility of performing a more detailed analysis of reaction mechanisms.

Most methods to compute reaction rate constants require the definition of a Collective Variable (CV), either to define the states of the system, its free energy, or to use it as a one dimensional Reaction Coordinate (RC) indexing the progress of the transition. In many situations, reactions go through one or a few channels in phase space. CVs should describe these channels with a minimal number of dimensions. Usually, CVs are defined thanks to chemical intuition or through the expert knowledge of the chemical system. They are typically based on key distances or angles associated with atoms central to the reaction mechanism. Nonetheless, this kind of heuristic approach can have some limitations especially when the studied mechanism is \textit{a priori} unknown. Automatic and data based approaches using various Machine Learning (ML) methods currently offer very appealing perspectives in this context. Recent reviews\cite{Glielmo2021, Chen2021, Gkeka2020, Ferguson2017} provide an overview of current options to propose CVs and discuss their advantages and drawbacks. These methods bear the promise of more systematic and efficient ways to define CVs, albeit at the expense of interpretability compared to intuitive CVs such as angles or distances. Nonetheless, machine-learned CVs are becoming common practice in the field. For example, Support Vector Machine (SVM) models trained on a set of data generated by molecular dynamics were used for exploring the configurational transitions of model protein molecules\cite{Sultan2018}. In material sciences, the combination of SVM and AIMD was used for the mechanistic study of the diffusion of Al atoms on Al (100) surface\cite{Pozun2012}. To the best of our knowledge, SVM has not been used to explore more complex reactive events, such as chemical bond breaking/formation catalyzed by an oxide material's surface such as proposed in the present work. 

To benchmark an innovative methodology based on the Hill relation for exploring reaction mechanisms occurring on catalytic materials, we chose in this work a relevant case study: the reactivity of water on the (100) orientation of $\gamma-$alumina, a widely used support in heterogeneous catalysis applied to biomass conversion\cite{Christiansen2013, Larmier2016}. Comprehensive DFT based studies have revealed the versatile nature of active sites (Lewis Al and Bronsted Al-OH), their thermodynamic properties\cite{Hass1998, DIGNE2002, DIGNE2004, Wischert2012, Pigeon2022} and their kinetic ones (TS and activation barriers) by using predominantly hTST calculations \cite{Christiansen2013, Larmier2016, Lu2016, Pan2008}. As for the study of many chemical reactions, especially in catalysis, most of the reaction rate constants are computed within the TST framework\cite{Piccini2022}. Unbiased AIMD simulations have been also applied to decipher the gamma-alumina's reactivity, its local structure and spectroscopic features, in the presence of liquid water in order to obtain a better understanding of phenomena occurring during the catalyst preparation or catalytic reaction \cite{NgouanaWakou2017, Reocreux2017}. TPS was used in particular for studying the catalytic reactivity of other oxide materials\cite{Lo2005, Bucko2009}, also in combination with the blue-moon ensemble formalism\cite{Rey2020}.

Methods based on the Hill relation and rare event simulation methods are rarely used for studying chemical reactions \cite{Roet2021} and to the best of our knowledge, they have never been used to describe reactions in heterogeneous catalysis. In particular, the AMS method has only been used for molecular dynamics applications to study the isomerization of small biomolecules\cite{Lopes2019} or a protein-ligand dissociation\cite{Teo2016}, up to now.

Hence, the aim of the present work is to highlight how AMS applied to AIMD rare event sampling, combined with ML approach, is able to compute reaction rate constants via the Hill relation in a relevant case study for heterogeneous catalysis. The CVs and RCs are built using SVM or Path Collective Variables and well-chosen chemical descriptors\cite{Branduardi2007}.

Considering the challenge of the chemical reactivity of the alumina catalysts highlighted before, we will aim at determining rate constants for a reaction network involving various water rotation, dissociation and association events on the (100) $\gamma$-alumina surface. 

This article is organized as follows. In the methods section, the general computational approach following a flowchart leading to the determination of rate constants is described. First, it is presented how the implementation of AMS coupled to a reference plane wave-DFT software enables the determination of rate constants. In a second part, numerical tools such as SVM and path collective variables (PCV) used to define CVs and RCs are presented. The results section first describes the catalytic model system of water activated on the $\gamma$-alumina surface, used to probe the theoretical approach. The constructions of CVs and RCs corresponding to the water molecule transformation path are then explained. Finally, the numerical values of reaction rates and the reactive trajectories are analyzed and compared with the standard hTST approach.

\section{Methods}
\label{section_methods}

\begin{figure}
	\centering
\includegraphics[width=14cm]{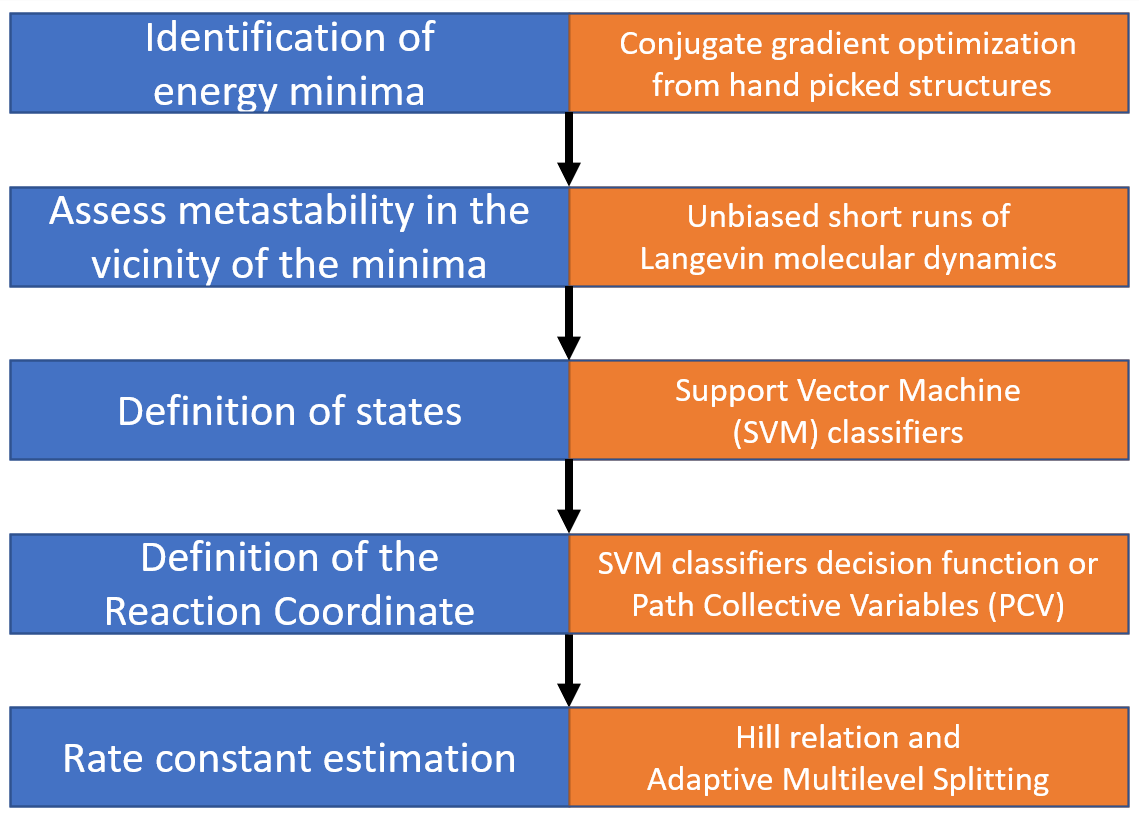}
	\caption{Global workflow to compute reaction rate constants With the Hill relation using Adaptive Multilevel Splitting and Machine Learning.}
	\label{Flowchart_global}
\end{figure}

The general flowchart of our approach is given in Figure~\ref{Flowchart_global}. The first step is the definition of states defined as the ensemble of structures in the vicinity of a local potential energy minimum characterizing either a reactant or a product. In practice, these configurations are sampled by running a short AIMD starting from minima identified on the Potential Energy Surface (PES). Then, using this trajectory, the function numerically defining states is obtained by SVM and well-chosen chemical descriptors. Depending on the reaction rate constant to compute, each state has to be labeled as reactant or product. A reaction coordinate (RC) is then built, for instance by using the decision functions of the classifiers previously used to define states. Once states and a RC are defined, AMS is run to obtain an estimate of the reaction rate constant of the Langevin dynamics which is assumed to model accurately the system dynamics. 

\subsection{Reaction rate constant estimation using AMS}

\paragraph{Motivation.} To compute rate constants of rare events by using the Hill relation~\eqref{Hill_relation}, the flux of trajectories leaving the initial reactant state $R$ (or the frequency at which trajectories leave $R$) must be evaluated. If the reactant state is properly defined, this quantity can be computed in a reasonably short time by unbiased MD. The difficulty lies in the estimation of the probability that a trajectory leaving $R$ is reactive (i.e. goes to a product state $P$), since the probability $p_{R \rightarrow P}(\partial R)$ is in most cases exceedingly small. The AMS algorithm is specifically designed to evaluate low probability events\cite{Cerou2007}. The key point of AMS is to propose a method that has a good behavior in terms of variance and computational efficiency to compute the probability $p_{R \rightarrow P}(\partial R)$. This is achieved by first decomposing the rare event of interest into a succession of less unlikely events, the target probability to estimate being the product of the conditional probabilities associated with the sub-events (see SI~Section~1). Moreover, the sub-events are built such that the associated conditional probabilities are all the same. This is indeed a desirable feature in order to reduce the overall variance of the estimator\cite{Lopes2019}. The mathematical analysis of the variance of the AMS estimator is provided in Ref.~\citenum{Cerou2007} and~\citenum{Cerou2019}. We focus here on the presentation of the algorithm adapted to MD rare events and only mention that this algorithm is unbiased\cite{CharlesEdouard2015UnbiasednessOS}. This means that, whatever the choice of the reaction coordinate $\xi$ and the number of replicas of the system (see below), repeating the algorithm sufficiently many times will always provide the same result in average, and this average value coincide with the target probability. On the other hand, the variance of the probability estimator depends on the quality of~$\xi$. This opens a way to define an iterative procedure to improve the definition of reaction coordinates, using the sampled reactive trajectories to define better reaction coordinates. 

\begin{figure}
	\centering
	\includegraphics[width=10cm]{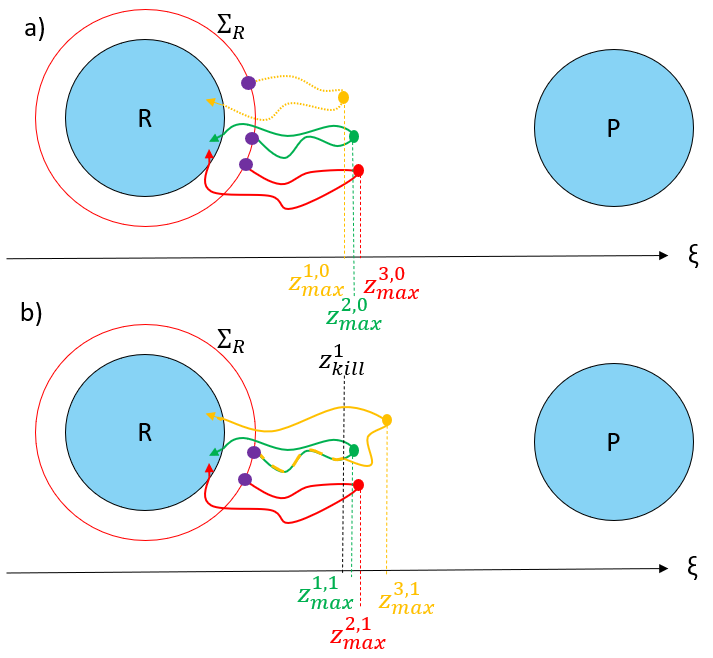}
	\caption{First iteration of the AMS algorithm with $k_{\mathrm{min}}=1$ and $N_{\mathrm{rep}} = 3$. Purple points represent the initial conditions on $\Sigma_R$. \textbf{a)} Identify the kill level $z_{\mathrm{kill}}^{1} = z_{\mathrm{max}}^{k_{\mathrm{min}},0}$ and kill the replicas such that $z_{\mathrm{max}}^{i,0} \leq z_{\mathrm{kill}}^{1}$, i.e. the orange replica.  \textbf{b)} Replace the killed replicas by the trajectory of one of the remaining replicas (the green one in this example) until the level $ z_{\mathrm{kill}}^{1}$ and continue the trajectory of the replica until it reaches either the state $R$ or the state $P$.} 
	\label{AMS_loop}
\end{figure}

\paragraph{Computing the flux and sampling initial conditions.} A separating surface $\Sigma_R$ close to $R$ is introduced for the estimation of the flux $\Phi_R$, to determine actual exits out of $R$\cite{Baudel2020}. This surface has to enclose the reactant state, so that any trajectory going from $R$ to $P$ has to cross $\Sigma_R$ (see Fig.~\ref{AMS_loop}). Indeed, the location of this surface allows to select the trajectories that make actual excursions off the state $R$, in contrast to trajectories that would only wander out of $R$ for a few steps and go back inside $R$ right away. The flux $\Phi_R$ is then evaluated by starting a dynamics in the state $R$, counting the number of times $n_{\mathrm{loop}-R\Sigma_R R }$ it goes from $R$ to $\Sigma_R$, crosses $\Sigma_R$ and goes back andto $R$, dividing this number by the overall time $t_{\mathrm{tot}}$: 
\begin{equation}
\label{flux_approx}
\Phi_R = \frac{n_{\mathrm{loop}-R\Sigma_R R }}{t_{\mathrm{tot}}} = \frac{1}{t_{\mathrm{loop}-R\Sigma_R R }},
\end{equation}
where $t_{\mathrm{loop}-R\Sigma_R R }$ is the average time that a trajectory takes to go out of $R$, cross $\Sigma_R$ and go back to $R$. Now that the calculation of the first term $\Phi_R$ has been discussed, let us focus on the second one: $p_{R \rightarrow P}(\partial R)$. Computation of the flux $\Phi_R$ simultaneously allows to generate some positions on the surface $\Sigma_R$, which will serve to estimate the probability $p_{R \rightarrow P}(\Sigma_R)$. Indeed, the estimated quantity is $p_{R \rightarrow P}(\Sigma_R)$ instead of $p_{R \rightarrow P}(\partial R)$, this does not bias the result as far as $R$ and $\Sigma_R$ are within the same metastable basin~\cite{Baudel2020}. These initial conditions must correspond to the first time a trajectory leaving $R$ reaches the level $\Sigma_R$. As efficient calculation of the flux and the sampling of initial conditions relies on parallelization strategies, a Fleming--Viot particle process is used in our implementation of this initialization procedure\cite{Binder2015}. The particles undergo independent molecular dynamics which means they can be run in parallel without requiring frequent communications. 

\paragraph{AMS Requirements.} To run an AMS estimation, the reactant state $R$ and the product state $P$ have to be defined. The surface $\Sigma_R$ has to be placed such that each trajectory linking the reactant and the product state goes through $\Sigma_R$. Its distance to the boundary of $R$ should be sufficiently small so that the sampling of initial conditions and the determination of the flux $\Phi_R$ (see the previous paragraph) is not exceedingly expensive in terms of computational cost. A number of replicas $N_\mathrm{rep}$ (or walkers) has to be defined, as well as a minimum number $k_\mathrm{min}$ of replicas to kill at each iteration of AMS. $N_\mathrm{rep}$ different initial conditions on the surface $\Sigma_R$ are selected uniformly among the initial conditions sampled following the procedure described in the previous paragraph (purple points in Figure~\ref{AMS_loop}). Finally a reaction coordinate $\xi$ should be defined to index the progression along the $R \rightarrow P$ transition. It has to be consistent with the states $R$ and $P$ which can be generally enforced by setting $\xi(\textbf{q}) = - \infty$ for $\textbf{q} \in R$ and $\xi(\textbf{q}) = + \infty$ for $\textbf{q} \in P$. 

\paragraph{AMS initialization.} First, all the replicas are run from their initial conditions on $\Sigma_R$ until either the $R$ or $P$ state is reached (see Fig.~\ref{AMS_loop}~\textbf{a)} which depicts an initialised set of three replicas). They are then iteratively updated until they all finish in the product state $P$. An illustration of an iteration is provided in Figure~\ref{AMS_loop}, the process being detailed in the next paragraph. In what follows, $\textbf{q}_t^{i,n}$ denotes the position of the $i$-th replica at time $t$ and iteration $n$. In particular, $\lbrace \textbf{q}_0^{i,n} \rbrace_{1 \leq i \leq N_{\mathrm{rep}}}$ are initial conditions on $\Sigma_R$. The method to estimate the probability is also summarized in the pseudo-code presented in SI~Section~1. 

\paragraph{AMS iteration.} Each iteration of the main AMS loop starts by defining the largest value of the RC for each replica at the $n$-th iteration as $z_{\mathrm{max}}^{i,n} = \underset{t}{\sup}(\xi (\textbf{q}_t^{i,n}))$. The replicas are then reordered by increasing values $z_{\mathrm{max}}^{i,n}$ (see Figure~\ref{AMS_loop}). According to the value of $k_{\mathrm{min}}$, the level at which positions are killed is identified as an empirical quantile: $z_{\mathrm{kill}}^{n+1} = z_{\mathrm{max}}^{k_{\mathrm{min}},n}$. This means that all the trajectories for which $z_{\mathrm{max}}^{i,n} \leq z_{\mathrm{kill}}^{n+1}$ are killed. The number of killed trajectories at this iteration is denoted by $\eta_{\mathrm{killed}}^{n+1}$. Note that $\eta_{\mathrm{killed}}^{n+1} \geq k_\mathrm{min}$ by construction, but it could happen that $\eta_{\mathrm{killed}}^{n+1} \geq k_\mathrm{min} + 1$ when several trajectories reach exactly the same $z_{\mathrm{max}}^{k_{\mathrm{min}},n}$. To keep the number of replica constant, $\eta_{\mathrm{killed}}^{n+1}$ trajectories have to be created by randomly branching $\eta_{\mathrm{killed}}^{n+1}$ trajectories among the remaining ones. More precisely, trajectories are duplicated until the first time they reach the level $z_{\mathrm{kill}}^{n+1}$ and then the dynamics is ran from these points until it reaches $R$ or $P$. In  fact, at each iteration, the estimated probability $1 - \frac{\eta_{\mathrm{killed}}^{n+1}}{N_{\mathrm{rep}}}$ is the probability for a trajectory to reach the surface $\Sigma_{z_{\mathrm{kill}}^{n+1}}$ starting on the surface $\Sigma_{z_{\mathrm{kill}}^{n}}$.

\begin{figure}
	\centering
	\includegraphics[width=12cm]{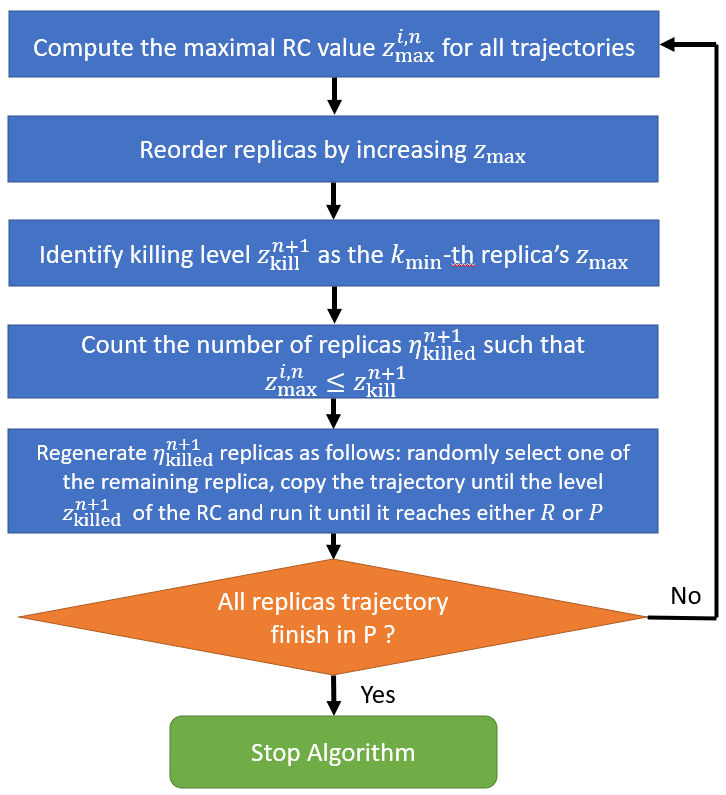}
	\caption{Flowchart of one iteration of AMS. $n$ is an iteration index and $i$ a replica index.} 
	\label{AMS_iteration_flowchart}
\end{figure}

Any AMS iteration can be summarized by the succession of the steps illustrated in Figure~\ref{AMS_iteration_flowchart}. 

\paragraph{AMS termination and probability estimator.} The AMS algorithm can terminate in two different manners. First, after a certain number of iterations, all the replicas reach the state $P$. In such a case, $N_{\mathrm{rep}}$ different reactive trajectories are obtained and the estimated transition probability is computed via:
\begin{equation}
\label{Proba_estimator_AMS}
\widehat{p}_{R \rightarrow P}(\Sigma_R) = \prod_{n = 1}^{n_\mathrm{max}} \left(1 - \frac{\eta_{\mathrm{killed}}^{n}}{N_{\mathrm{rep}}} \right),
\end{equation} 
where $n_\mathrm{max}$ is the final number of iterations of the algorithm. The second option (not explicitely presented on Figure~\ref{AMS_iteration_flowchart} since the RC is typically chosen so that this does not happen) is that at a certain iteration $n$, $\eta_{\mathrm{killed}}^n$ is equal to the total number of replicas of the algorithm. This can happen if at some point all the copied replicas have the same value of $z_{\mathrm{max}}^{i,n}$. This termination event is called "failure" as the algorithm is not able to provide reactive trajectories and the estimated probability is $\widehat{p}_{R \rightarrow P}(\Sigma_R) = 0$, consistently with expression~\eqref{Proba_estimator_AMS}. Such a situation can be encountered if the system is stuck and all the replicas are progressively replaced by the copy of a single replica. It is also possible that the replicas reach their maximum in $\xi$ in a zone of the phase space on which the reaction coordinate $\xi$ remains constant while the trajectories are different.

It is possible to estimate the statistical error on the estimated probability $\widehat{p}_{R \rightarrow P}(\Sigma_R)$ in~\eqref{Proba_estimator_AMS} by repeating the estimation of the probability $M_\mathrm{real}$ times. These realisations should be independent and can take advantage of parallel architecture of current super-computers. The confidence intervals presented in the results section all correspond to a 90\% confidence. More details can be found in SI~Section~2.

\paragraph{Multiple states case.} Defining state $R$ and $P$ in multiple state case needs a specific treatment to compute state to state reaction rates. Two main approaches are proposed and detailed in SI~Section~3. The first one samples all possible trajectories starting from a given state. The second approach more specifically focuses on the targeted transition. An illustration and comparison of the two approaches are provided in the results section. 

\paragraph{Implementation with a plane wave DFT code.} The AMS algorithm and the sampling of initial conditions was implemented in Python scripts calling the VASP software for AIMD simulations\cite{Kresse1993, Kresse1999}. All DFT simulations parameters are listed in SI~Section~4.1 while AIMD parameters are presented in SI~Section~4.2. Some slight modifications have been implemented in the VASP code to allow for different stopping conditions of the VASP MD runs. More details concerning the implementation can be found in the SI~Section~5. The various repetitions $M_\mathrm{real}$ of the AMS estimation can be run independently in parallel. The Fleming--Viot particle scheme also allows for in dependant runs, communications are required only infrequently allowing arbitrary number of particles ran independently in parallel. The development of the scripts and the testing was mostly done on \textit{ENER440} calculator at IFPEN. Results presented in the following section come from simulations ran on \textit{Joliot-Curie}(Genci) and \textit{Topaze}(CCRT).

\subsection{Tools to define states and reaction coordinates}

Let us conclude this section describing the methods by introducing useful tools that will be used to define the states and the reaction coordinates in the next section.

\paragraph{Representation of chemical structures.} Reaction coordinates and states definitions must be invariant under rotation, translation and symmetries of the system as well as by permutation of identical atoms. Since description relying on Cartesian coordinates do not exhibit these properties, substantial work was conducted to find representations of atomic systems invariant by Galilean transformations and other symmetries, in particular in the field of ML empirical potentials\cite{Behler2007, Bartok2013, Drautz2019, Chen2020}. We chose the smooth overlap of atomic positions (SOAP)\cite{Bartok2013} descriptor allowing to capture enough information on atomic environments to reach errors of the order of 1 meV for potential energy surface fitting\cite{BartokPartay2010, Bartok2018}. This descriptor turned out to be sufficient for our needs as illustrated in the result section. The detailed parameters used to compute SOAP descriptor using \textit{dscribe} Python package\cite{Himanen2020} can be found in SI~Section~4.3. 

\paragraph{Support Vector Machine.}  A linear SVM model is designed to find the highest margin separation plane between two sets of labeled points. The margin denotes the minimal distance between the plane and the labeled points. The details concerning this optimization problem can be found in ML textbooks\cite{Murphy2022} or the \textit{scikit-learn} documentation\cite{Pedregosa2012}. The important result for this work is that, once the optimization problem is solved, only a certain subset of the total training set is used in the definition of the plane. These are the so-called support vectors which are the closest to the separation plane. The vector normal to this plane and the scalar defining its position is thus a linear combination of the support vectors. The classifier decision function is the algebraic distance to the plane multiplied by a scaling factor chosen so that the decision function value on support vectors which are not outliers is either $1$ or $-1$. To define multiple states using SVM, the one versus all approach was chosen, as made precise later on in the result section dedicated to the definition of states. Linear SVM models were trained using the SVC routine of \textit{scikit-learn} package with a linear kernel\cite{Pedregosa2012}. The data were normalized using the standard scaler implemented in the same package. The regularization parameter was kept to the default value~$1$ as, after cross validation, the classification scores on the test sets were always 100\%.

\paragraph{Path Collective Variables (PCV).} The principle of PCVs is to first define a reference path for the transition as a sequence of structures $\left\{ \textbf{R}_i\right\}_{0 \leq i \leq L-1}$. These structures are represented with a numerical descriptor, here the SOAP descriptor. A reaction coordinate is then constructed as\cite{Branduardi2007}:
\begin{equation}
\label{PCV_formula}
s(\textbf{R}) = \frac{\sum_{i=0}^{L-1} i \; \mathrm{e}^{- \lambda d(\textbf{R}_i, \textbf{R})} }{\sum_{i=0}^{L-1} \mathrm{e}^{- \lambda d(\textbf{R}_i, \textbf{R})}},
\end{equation}
where $d$ is a distance here the Euclidean norm. The parameter $\lambda$ has to be of the order of variation of the inverse distances between two consecutive structures along the path. If the structures along the path are not evenly spaced along the path according to the distance $d$, a sequence of values $\lambda_i$ can be used instead. In the present case, we chose $\lambda_i$ as:
\begin{equation}
    \left\{ 
    \begin{aligned}
    &\lambda_i^{-1} = \frac{1}{2}\left( d(\textbf{R}_{i-1}, \textbf{R}_i) + d(\textbf{R}_i, \textbf{R}_{i+1}\right)) \\
    & \lambda_0^{-1} = d(\textbf{R}_{0}, \textbf{R}_1) ; ~ \lambda_{L-1}^{-1} = d(\textbf{R}_{L-2}, \textbf{R}_{L-1}).
    \end{aligned}
    \right. 
\end{equation}
PCVs were directly implemented in the python scripts used for the reaction coordinate evaluation during the dynamics. 

\section{Results and discussion}

\subsection{$\gamma$-\ce{Al2O3} models and definition of states}

\begin{figure}
	\centering
	\includegraphics[width=12cm]{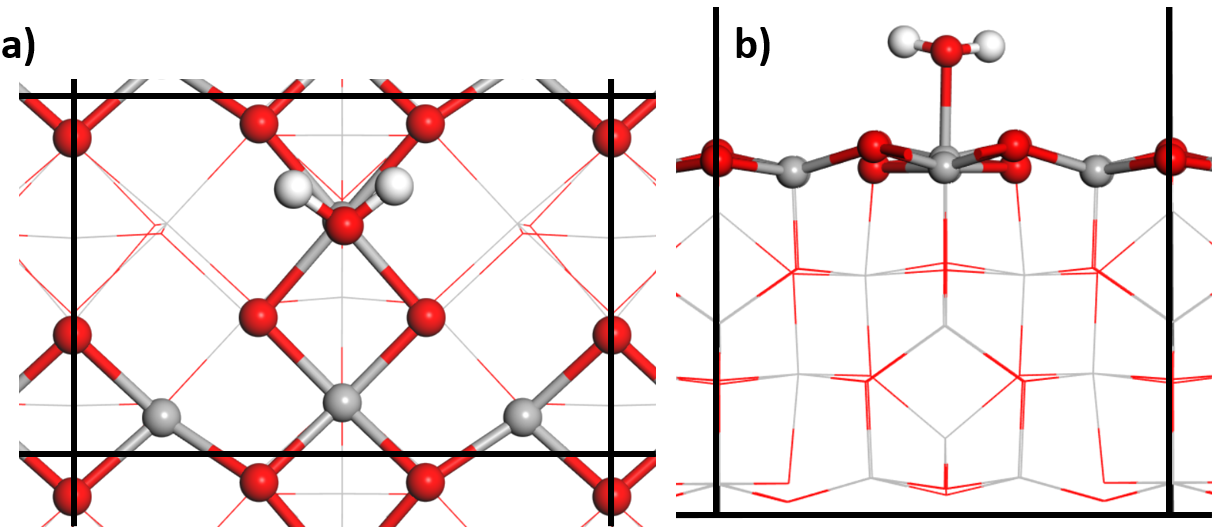}
	\caption{Representation of one water molecule adsorbed on an aluminum site of the $(100)$-$\gamma$-alumina surface model. Surface atoms are represented as ball and sticks while subsurface ones are represented as lines. Colors: Red: Oxygen, Grey: Aluminum, White: Hydrogen, Black: limit of the periodic cell. \textbf{a)} Top view; \textbf{b)} Side view.}
	\label{alumina}
\end{figure}

\paragraph{Model of the catalytic system.} The catalytic case study chosen to benchmark the previously presented method is the transformation of a water molecule adsorbed on the (100) $\gamma$-alumina surface\cite{DIGNE2002, DIGNE2004, Pigeon2022}. A representation of the $\gamma$-\ce{Al2O3} surface on which one water molecule is chemically adsorbed without dissociation on an aluminum Lewis site is given in Figure~\ref{alumina}. More information about the alumina slab used is provided in SI~Section~4.1.

The first step is to identify various potential energy minima corresponding to the metastable states of the water molecule adsorbed on the surface either in a dissociative mode or non-dissociative modes. As described in what follows, the dissociative modes lead to the formation of two hydroxyl (OH) groups : the first one is formed upon the transfer of a H atom of the water molecule to an O site of the surface; the second one results from the native water molecule. This systematic exploration confirms previous DFT studies where the minima were identified by running multiple geometry optimizations starting from various initial conditions\cite{DIGNE2002, DIGNE2004}. 

\begin{figure}
	\centering
	\includegraphics[width=12cm]{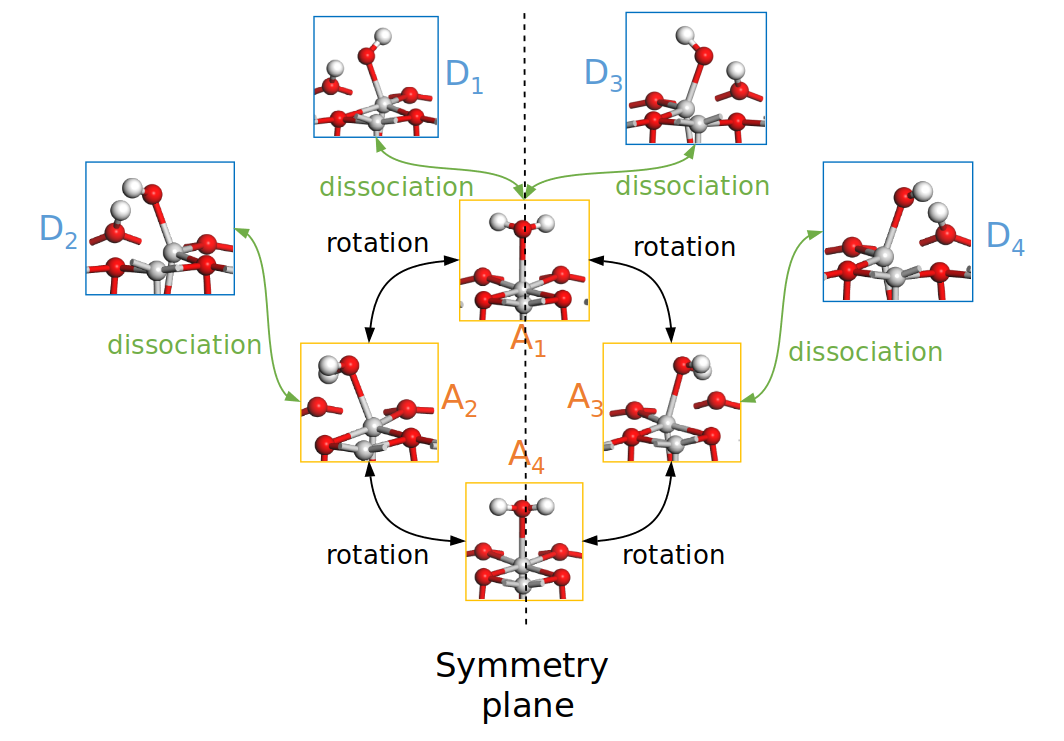}
	\caption{Representation of the main different minimum energy structures corresponding to metastable states for the water molecule adsorbed on the (100) $\gamma-\mathrm{Al}_2 \mathrm{O}_3$ surface. Arrows represent transitions that might occur. Color legend: gray: aluminum, red; oxygen, white: hydrogen}
	\label{states_Ai_Di}
\end{figure}

\paragraph{Data set generation to learn states.} Once local minima are identified, the metastability of the basins surrounding them should be assessed because these local minima should be sufficiently separated from other local minima. To quantify this, two AIMD trajectories of 1~ps each were run starting from each minimum. The first AIMD was run with a friction parameter of $5~\mathrm{ps}^{-1}$ to thermalize the system faster, while the second one was run with $\gamma=0.5~\mathrm{ps}^{-1}$. If the system ends up in another potential energy well during this second part of the trajectory, then the initial well is not considered relevant to be qualified as a metastable state. At 300 K, multiple transitions between all basins were observed, thus all the potential wells cannot be considered metastable and relevant so as to mimic realistic chemical reactions. At 200 K, 8 genuine metastable states could be identified denoting that at this temperature, the system better mimic chemical reaction conditions. The various identified states are named $A_i$ or $D_i$ depending on whether the state corresponds to a non-dissociated adsorbed water molecule or to two surface hydroxyls after water dissociation, respectively (see Fig.~\ref{states_Ai_Di}). Some of these states are in fact identical as there exists a plane symmetry in this structure and thus these metastable potential energy wells should be gathered in the same state. For example, the wells $D_1$ and $D_3$ are symmetrically identical.

The numerical definition the states $A_{1}, A_{2}A_3, A_{4}, D_1D_3, D_{2}D_{4}$ were built using one versus all (1-vs-all) linear SVM classifiers decision function $f_{X\mathrm{-vs-all}}$. For instance, the state $A_1$ is defined as $\left\{ \textbf{q} \ \middle| f_{A_{1}\mathrm{-vs-all}}(SOAP(\textbf{q})) \leq -1 \right\}$. To train these models, the data used was a 1~ps MD trajectory at 50 K starting from each local minimum. The point of running a MD at a lower temperature was to obtain points close to the minimum of the potential energy well. The dynamics starting was run with a friction parameter of $5~\mathrm{ps}^{-1}$ during $1~\mathrm{ps}$ for equilibration, then run during $1~\mathrm{ps}$ with $\gamma=0.5~\mathrm{ps}^{-1}$. 

The production runs of these trajectories were used to train the SVM classifiers. Only one SOAP descriptor centered on the oxygen atom of the adsorbed water molecule was used as features in the training set. With the parameters mentioned in the SI~Section~4.4, this leads to an array of size 2100 to describe each structure. Before training the model, the variation of each dimension of the SOAP descriptors were scaled to have zero mean and unit variance. The test score of the SVM model was 100\% in every case, which indicates that the set of structures represented with SOAP descriptors are linearly separable.  On the other hand, trying to separate the SOAP descriptor of the trajectories starting from two symmetric minima such as $D_1$ and $D_3$ systematically led to smaller test scores. This indicates that the well surrounding these two minima are indeed similar in the sense of the SOAP descriptor. 

\begin{figure}
	\centering
	\includegraphics[width=10cm]{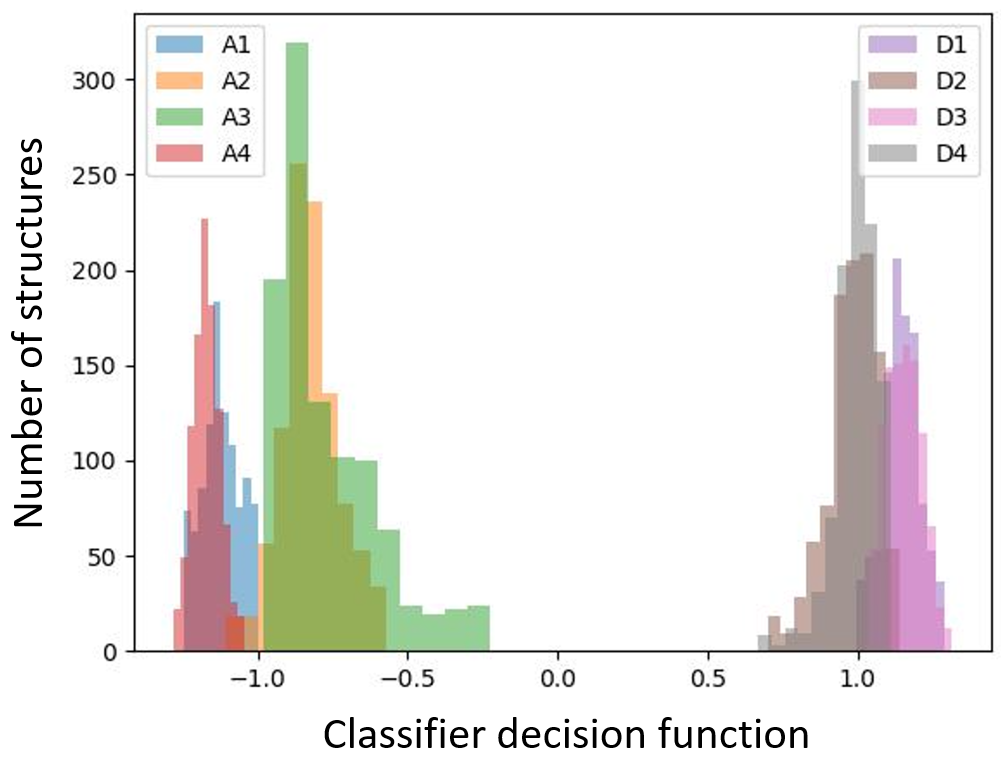}
	\caption{Histogram of $A_1$-vs-$D_1$ SOAP-SVM CV on the whole labelled dataset.}
	\label{SVM_histogram}
\end{figure}

As a side remark, in a situation where symmetries of states are unknown, using this kind of approach can help to identify some similarities. In Figure~\ref{SVM_histogram}, an histogram of the decision function of a $A_1$-vs-$D_1$ SOAP-SVM classifier is plotted. The various colors represent the different labelled states. It is clear that this CV allows to differentiate the $A$ and $D$ states. Moreover, according to this criterion, the $A_2$ and $A_3$ as well as well as the $D_1$/$D_3$ and $D_2$/$D_4$ groups of points bear some similarities for reason of symmetry. 

\paragraph{Definition of reaction coordinates (RCs).} The first RCs used to perform AMS simulations are the various 1-vs-all SVM decision functions. These RCs are therefore named "1-vs-all SOAP-SVM RC" in the following sections. Some more specific RCs are built using the same approach targeting a specific transition from a state to another. In this case, the decision function is obtained by separating only the two targeted states. The corresponding RCs are termed "1-vs-1 SOAP-SVM RC". Finally, a Path Collective Variables (PCV), termed "SOAP-PCV" is also used as reaction coordinate to index the progression of AMS replicas. The SOAP-PCV RCs differ depending on the reference path. We consider here paths built by an interpolation of the z matrix representations of the minima of two metastable basins\cite{Optnpath}. The associated RCs are termed "interpolated SOAP-PCV". 

\subsection{Analysis of AMS rate constants}

In this section, we analyze first the sensitivity of the reaction rates to two key parameters, the number of replicas ($N_\mathrm{rep}$) and the number of repetitions of the probability estimation ($M_\mathrm{real}$). These parameters also govern the computational cost and how this cost can be distributed on multiple CPUs, taking advantage of the parallel architecture of current supercomputers. Then, the other impacting choices on the precision of the reaction rate constant are the RC and the states, also investigated in what follows. The reaction rate constants obtained for each observed transition are finally compared to values computed from hTST. 

\paragraph{Parallel calculations against precision.} The effect of the number of replicas ($N_\mathrm{rep}$) and the number of AMS repetitions ($M_\mathrm{real}$) is evaluated for a fixed number of initial conditions $N_\mathrm{rep} M_\mathrm{real}$, which roughly corresponds to a fixed computational cost. Indeed, assuming that every branching during one AMS realization has the same cost in average and that $\eta_\mathrm{killed}^n$ is constant and equals $k_\mathrm{min}$ at all steps of the AMS realization, the cost of one AMS realisation is given by the product of the number of AMS iterations ($n_\mathrm{max}$) and the number of killed replicas ($k_\mathrm{min}$). Under these assumptions, the AMS estimator~\eqref{Proba_estimator_AMS} writes:
\begin{equation}
\label{AMs_ideal_estimator}
\widehat{p} = \left(1 - \frac{k_\mathrm{min}}{N_\mathrm{rep}} \right)^{n_\mathrm{max}}.
\end{equation}
Assuming that $\frac{k_\mathrm{min}}{N_\mathrm{rep}}$ is small, the computational cost of a single AMS simulation is:
\begin{equation}
\label{AMS_approx_cost}
k_\mathrm{min} n_\mathrm{max} \approx -N_\mathrm{rep} \ln\left( \widehat{p}\right).
\end{equation}
Taking into account number of repetitions of the algorithm $M_\mathrm{real}$, the final cost of a reaction rate constant estimation is $- M_\mathrm{real} N_\mathrm{rep}\ln\left( \widehat{p}\right)$. Considering the current implementation of AMS, $M_\mathrm{real}$ realisation of AMS can be run in parallel. The objective is to find the minimal value of $N_\mathrm{rep}$ to better distribute the computational cost on multiple parallel realisations. With too few replicas, the intrinsic variance of the AMS estimator can be so large that the confidence interval of the estimated probability contains 0, leading to not interpretable results. Table~\ref{n_rep_table} reports the evolution of water rotation rate constants $k_{A_1 \rightarrow A_2 A_3}$ calculated with AMS for various values of $N_\mathrm{rep}$ and $M_\mathrm{real}$ by using the "$A_1$-vs-all-SOAP-SVM" reaction coordinate and states defined as $R~=~A_1$ and $P=~A_2A_3~\cup~A_4~\cup~D_1D_3~\cup~D_2D_4$.
\begin{table}
	\caption{Estimation of probability, rate and the corresponding accuracy at 90\% confidence for water rotation. The number of initial conditions $M_\mathrm{real}N_\mathrm{rep}$ was varying $M_\mathrm{real}$ and $N_\mathrm{rep}$.  $R~=~A_1$, $P=~A_2A_3~\cup~A_4~\cup~D_1D_3~\cup~D_2D_4$, $\xi$ = $A_1$-vs-all SOAP SVM RC.}
	\label{n_rep_table}
	\begin{tabular}{lllll}
		\hline
		$M_\mathrm{real}$ & $N_\mathrm{rep}$ & $t_{\mathrm{loop}-R\Sigma_{A1}R} $ (fs) & $p_{A_1 \rightarrow A_2A_3}(\Sigma_{A_1})$ & $k_{A_1 \rightarrow A_2 A_3}~(\mathrm{s}^{-1})$ \\
		\hline
		5  & 400 & $108 \pm 5$ & $(3.73 \pm 3.03) 10^{-3}$ & $(3.67 \pm 2.99) 10^{10}$  \\
		10 & 200 & $110 \pm 5$ & $(3.38 \pm 1.56) 10^{-3}$ & $(3.08 \pm 1.43) 10^{10}$  \\
		20 & 100 & $101 \pm 5$ & $(3.47 \pm 1.96) 10^{-3}$ & $(3.21 \pm 1.82) 10^{10}$  \\
		\hline
	\end{tabular}
\end{table}

By definition $t_{\mathrm{loop}}$ is not impacted by $N_\mathrm{rep}$ or $M_\mathrm{real}$. The target value of probability and rate are little impacted in the present case, which is not the case for the variance. The choice of $N_\mathrm{rep} = 200$ and $M_\mathrm{real}=10$ is sufficient to obtain a $~A_1$ to $~A_2A_3$ water rotation rate of $3.1~10^{10}~\mathrm{s}^{-1}$ with the 90\% confidence interval of $\left[1.65~10^{10}\mathrm{s}^{-1},~4.51~10^{10}\mathrm{s}^{-1}\right] $. Similar precision can be obtained with $N_\mathrm{rep} = 100$ and $M_\mathrm{real}=20$. Therefore, it is important to perform the AMS simulations a certain number of times ($M_{real}$) in order have a proper variance estimation. Hence, for a similar computational cost in CPU time, satisfying accuracy can be obtained using $M_\mathrm{real} \geq 10$. 

\paragraph{Impact of the definition of reaction coordinates and states.} The definitions of the states $R$ and $P$ determine the type of trajectories that can be sampled by the algorithm. The choice of the reaction coordinate impacts the quality of this sampling. For instance, exploring all types of trajectories from $A_1$ to any other states, requires to sample initial conditions on $\Sigma_{A_1}$, set $R~=~A_1$ and $P=~A_2A_3~\cup~A_4~\cup~D_1D_3~\cup~D_2D_4$. Using the $A_1$-vs-All SOAP-SVM RC to sample trajectories ending in $P$ leads to the results presented in Table~\ref{A1-vs-all}. 
\begin{table}
    \caption{Transition rates leaving $A_1$ estimated using $A_1$-vs-all SOAP-SVM RC,              $N_\mathrm{rep} = 200$, $M_\mathrm{real} = 10$, $R~=~A_1$ and                               $P=~A_2A_3~\cup~A_4~\cup~D_1D_3~\cup~D_2D_4$. As the results come from the same AMS $t_{\mathrm{loop}-R\Sigma_{R}R}$ is constant and equal to $110 \pm 5$ fs.}
    \label{A1-vs-all}
    \begin{tabular}{lll}
    \hline
    Transition &  $p_\mathrm{Transition}(\Sigma_{A_1})$ &              $k_\mathrm{Transition}~(\mathrm{s}^{-1})$   \\
    \hline
    $A_1 \rightarrow A_2A_3$   & $\left(3.38\pm 1.56\right) \, 10^{-3}$ &         $\left(3.17 \pm 1.43\right) \, 10^{10} $ \\
    $A_1 \rightarrow D_1D_3$   & $\left(1.79\pm 1.86\right) \, 10^{-3}$ &         $\left(1.63 \pm 1.70\right) \, 10^{10} $ \\
    $A_1 \rightarrow A_4$      & $\left(3.66\pm 6.02\right) \, 10^{-7}$ &         $\left(3.44 \pm 5.50\right) \, 10^{6} $ \\
    \hline
    \end{tabular}
\end{table}
This approach allows to sample the transition $A_1 \rightarrow A_2A_3$ with a reasonable accuracy according to the estimate of the rate constant's variance. However, the less probable transitions ($A_1 \rightarrow D_1D_3$ and $A_1 \rightarrow A_4$) are under-sampled and the rate estimations are not precise enough as the 90\% confidence interval contains 0. Moreover, the direct transition from $A_1$ to $D_2 D_4$ being so rare that it has not even been sampled. To more accurately quantify the transition $A_1 \rightarrow D_1D_3$, more specific RCs must be used. The results obtained with two other RCs are compared in Table~\ref{cv_table}. Changing the reaction coordinate $A_1$-vs-all SOAP-SVM into $A_1$-vs-$D_1$ SOAP-SVM for AMS does not significantly improve the rate constant precision as the estimated variance is still so large that 0 is contained in the confidence interval. This is due to the fact that in view of the definition of $R$ and $P$, AMS still samples trajectories that are of no interest such as the rotation $A_1 \rightarrow A_2A_3$. To observe only $A_1 \rightarrow D_1D_3$ reactive trajectories, one possibility would be to set $R~=~A_1$ and $P~=~D_1D_3$. However, as the AMS iteration stops once the trajectories finish either in $R$ or $P$, a trajectory including the $A_1 \rightarrow A_2A_3$ rotation would consume too much computational time before going to $R$ or $P$ as the state $A_2A_3$ is metastable. Hence $R$ and $P$ must be defined differently. Considering transition starting from $A_1$, with the choice $R~=~A_1~\cup~A_2A_3~\cup~A_4~\cup~D_2D_4$, $P=D_1D_3$, and initial conditions sampled on $\Sigma_{A_1}$, AMS is compelled to sample $A_1 \rightarrow D_1D_3$ trajectories. The difference here with the previous case is that, if a rotation $A_1 \rightarrow A_2A_3$ is observed in the course of the algorithm, then it will be stopped once it enters the $A_2A_3$ state and be considered as a non reactive trajectory. Such trajectories will ultimately be discarded and replaced by trajectories having higher values $z_\mathrm{max}$ of the chosen reaction coordinate defined in a way so as to enhance the sampling of trajectories between the desired metastable states. Both the quality of the reaction coordinate and the choice of the $R$ and $P$ states are important to obtain precise results for the $A_1 \rightarrow D_1D_3$ transition (see Table~\ref{cv_table}). In our case study, the necessity to change the definition of $R$ and $P$ might be due to the difference of the transition probability between the rotation $A_1 \rightarrow A_2A_3$ and the water dissociation $A_1 \rightarrow D_1D_3$. Indeed, the half size of confidence intervals is larger than the target rate in the case where any type of rotations can be sampled, while constraining the AMS to sample only $A_1 \rightarrow D_1 D_3$ trajectories leads to smaller confidence intervals. In the present case the interpolated SOAP-PCV RC is not significantly better that the $A_1$-vs-$D_1$-SOAP-SVM RC in term of variance as the 90\%  confidence error represents 97\% of the target values while for the $A_1$-vs-$D_1$-SOAP-SVM RC is is 89\%. 
\begin{table}
    \caption{Variation of the RC and reactant / product states $R$ and $P$ to sample the $A_1 \rightarrow D_1D_3$ transition with $N_\mathrm{rep} = 200$, $M_\mathrm{real} = 10$ and initial conditions sampled on $\Sigma_{A_1}$}
    \label{cv_table}
    \begin{tabular}{llll}
    \hline
    RC & $t_{\mathrm{loop}-R\Sigma_{A_1}R}$ (fs)& $p_{A_1 \rightarrow D_1D_3}(\Sigma_{A_1})$ & $k_{A_1 \rightarrow D_1D_3}~(\mathrm{s}^{-1})$   \\
    \hline
    \multicolumn{4}{l}{$R~=~A_1$ ; $P=~A_2A_3~\cup~A_4~\cup~D_1D_3~\cup~D_2D_4$ }\\
    \hline
    $A_1$-vs-all-SOAP-SVM    & $110 \pm 5$ & $\left(1.79\pm 1.86\right) \, 10^{-3}$ &         $\left(1.63 \pm 1.70\right) \, 10^{10} $ \\
    $A_1$-vs-$D_1$-SOAP-SVM  & $105 \pm 3$ & $\left(1.81\pm 1.98\right) \, 10^{-5}$ &         $\left(1.72 \pm 1.88\right) \, 10^{8} $ \\
    interpolated SOAP-PCV    & $104 \pm 4$ & $\left(1.95\pm 2.26\right) \, 10^{-4}$ &         $\left(1.87 \pm 2.17\right) \, 10^{9} $  \\
    \hline
    \multicolumn{4}{l}{$R~=~A_1~\cup~A_2A_3~\cup~A_4~\cup~D_2D_4$ ; $P=D_1D_3$}\\
    \hline
    $A_1$-vs-$D_1$-SOAP-SVM  & $105 \pm 2$ & $\left(3.31 \pm 2.97\right) \, 10^{-4}$ &         $\left(3.15 \pm 2.83\right) \, 10^{9} $  \\
    interpolated SOAP-PCV    & $108 \pm 2$ & $\left(1.78 \pm 1.73\right) \, 10^{-4}$ &         $\left(1.64 \pm 1.59\right) \, 10^{9} $  \\
    \hline
    \end{tabular}
\end{table}

\paragraph{Comparison of the rate constants calculated with AMS and with hTST.} Various rate constants involved in the reaction networks of Figure~\ref{states_Ai_Di} were computed using AMS. Various reaction coordinates and various definitions of the states $R$ and $P$ were used to obtain the results presented in Table~\ref{all_rates_table}. For the sake of clarity, the choice of $R$, $P$, RC and AMS parameters for each transition are listed in SI~Table~1. These rates obtained by AMS are directly compared to the reaction rate constants computed from the static hTST approach. Activation free energies calculated with hTST are reported in SI~Table~3. and they qualitatively compare with previously published DFT data.\cite{Pan2008} 

\begin{table}
    \caption{Transition rate constants for all the transitions observed in this study with 90\% confidence interval for AMS results.}
    \label{all_rates_table}
    \begin{tabular}{lll}
    \hline
    Transition & $k_\mathrm{Transition-AMS}~(\mathrm{s}^{-1})$ & $k_\mathrm{Transition-hTST}~(\mathrm{s}^{-1})$   \\
    \hline
    \multicolumn{3}{l}{Water rotations}\\
    \hline
    $A_1 \rightarrow A_2A_3$  & $\left(3.08 \pm 1.43\right) \, 10^{10} $ & $7.55 \; 10^{10}$ \\
    $A_2A_3 \rightarrow A_1$  & $\left(1.49 \pm 0.46\right) \, 10^{11} $ & $2.06 \; 10^{12}$ \\
    \hline
    $A_2A_3 \rightarrow A_4$  & $\left(4.33 \pm 2.20\right) \, 10^{10} $ & $3.64 \; 10^{10}$ \\
    $A_4 \rightarrow A_2A_3$  & $\left(2.35 \pm 0.87\right) \, 10^{11} $ & $5.66 \; 10^{11}$ \\
    \hline
    $A_1 \rightarrow A_4$  & $\left(3.34 \pm 6.56\right) \, 10^{6}  $ & $2.04 \; 10^{8}$  \\
    $A_4 \rightarrow A_1$  & $\left(1.34 \pm 0.68\right) \, 10^{10} $ & $8.65 \; 10^{10}$ \\
    \hline
    \multicolumn{3}{l}{Hydroxyl rotation}\\
    \hline
    $D_1D_3 \rightarrow D_2D_4$ & $\varnothing$ & $2.38 \; 10^{9}$ \\
    $D_2D_4 \rightarrow D_1D_3$ & $\left(2.86 \pm 4.71\right) \, 10^{8} $ & $4.15 \; 10^{9}$\\
    \hline
    \multicolumn{3}{l}{Formation and dissociation of water} \\
    \hline
    $A_1 \rightarrow D_1D_3$ & $\left(1.64 \pm 1.59\right) \, 10^{9}  $ & $3.37 \; 10^{11}$\\
    $D_1D_3 \rightarrow A_1$ & $\left(2.32 \pm 1.59\right) \, 10^{10} $ & $1.13 \; 10^{12}$\\
    \hline
    $A_2A_3 \rightarrow D_2D_4$ & $\left(7.86 \pm 7.53\right) \, 10^{9}  $ & $5.45 \; 10^{13}$\\
    $D_2D_4 \rightarrow A_2A_3$ & $\left(1.28 \pm 0.54\right) \, 10^{11} $ & $1.17 \; 10^{13}$\\
    \hline
    $A_2A_3 \rightarrow D_1D_3$ & $\varnothing$ & $\varnothing$ \\        
    $D_1D_3 \rightarrow A_2A_3$ & $\left(2.33 \pm 3.14\right) \, 10^{8} $ &  $\varnothing$\\
    \hline
    \end{tabular}\\
\end{table}

Reaction rate constants obtained by harmonic approximation are consistently higher than those obtained via the Hill relation and AMS for the Langevin dynamics, with one single exception for the $A_2A_3 \rightarrow A_4$ rotation. Assuming that the friction parameter is set so that Langevin dynamics reproduces accurately the system's dynamics, the AMS rate constants should be more precise than the TST ones due to the intrinsic overestimation of rates of TST as mentioned in the introduction. The harmonic approximation of the potential energy surface for fast approximations of free energies can lead to large errors. In particular, entropic effects are usually mistreated by hTST approaches as it was underlined by previous theoretical studies based on transition path sampling and blue moon ensemble simulations\cite{Rey2020} or other approaches\cite{Collinge2020}. In the present case, this might be the reason for the important overestimation of the rates of formation and dissociation events. Especially in the case of the $A_2A_3 \rightarrow D_2D_4$ transition, the approximation of the TS free energy is so bad that the activation free energy is negative (as reported in SI~Table~3) which leads to the large overestimation of the rates. 

Under the assumption of a correctly parameterized dynamics, the values presented in Table~\ref{all_rates_table} allow to realize that most water rotations are at least one order of magnitude faster than dissociation events. Only the direct $A_1 \rightarrow A_4$ rotation seems to occur less frequently. The formation of water happens on the same timescale as the fast water rotations depending on the hydroxyls conformation. This ordering has to be compared to the one from hTST rate constants. The quickest changes are the water formation and dissociation events. The slowest formation event occur as frequently as the fastest water rotation.

Using the presented approach to compute various reaction rate constants, especially those of forward and backward reactions, one can deduce also reaction free energies:
\begin{equation}
\label{eq_constant}
K_{R \rightarrow P} = \frac{k_{R \rightarrow P}}{k_{P \rightarrow R}},
\end{equation}
and
\begin{equation}
\label{reaction_heat}
\Delta G_{R \rightarrow P} (T) = -\mathcal{N}_\mathrm{A} k_{\mathrm{B}}T \ln\left(K_{R \rightarrow P}\right),
\end{equation}
where $\mathcal{N}_\mathrm{A}$ is the Avogadro number and $K_{R \rightarrow P}$ is the reaction equilibrium constant. 

\begin{table}
    \caption{Reaction heats at 200 K computed from Table~\ref{all_rates_table} and hTST} 
    \label{reaction_heat_tables}
    \begin{tabular}{lll}
    \hline
    & AMS Value $(\mathrm{kJ.mol}^{-1})$ & hTST Value $(\mathrm{kJ.mol}^{-1})$\\
    \hline
    \multicolumn{3}{l}{Water rotations} \\
    \hline
    $\Delta G_{A_1 \rightarrow A_2 A_3} $     & $2.62 \pm 2.66 $ & $5.50$\\
    $\Delta G_{A_2 A_4 \rightarrow A_4} $     & $2.81 \pm 2.83$ & $4.56$\\
    $\Delta G_{A_1 \rightarrow A_4} $         & $13.8 \pm 4.43$ & $10.1$\\
    \hline
    \multicolumn{3}{l}{Hydroxyl rotations} \\
    \hline
    $\Delta G_{D_1 D_3 \rightarrow D_2 D_4} $ & $\varnothing$ & $0.93$\\
    \hline
    \multicolumn{3}{l}{Water dissociations} \\
    \hline
    $\Delta G_{A_1 \rightarrow D_1 D_3} $     & $4.41 \pm 3.88$ & $-2.56$\\
    $\Delta G_{A_2 A_3 \rightarrow D_2 D_4} $ & $4.64 \pm 3.54$ & $2.01$\\
    \hline
    \end{tabular}
\end{table}

The values of the reaction free energies of Table~\ref{reaction_heat_tables} allow to identify that according to the harmonic approximation, the most stable state should be the $D_1 D_3$, while the most stable one identified with the AMS method is $A_1$ for T=200 K. Previous ab initio thermodynamic studies within harmonic approximations also identified that the dissociative state is more favored.\cite{DIGNE2002, DIGNE2004} Here also, one may suspect that entropic contributions may be at the origin of the change in the stability order. In particular, within the harmonic approximation, it is assumed that the adsorbed water molecule in $A_1$ state and in $D_1D_3$ has similar rotational and transnational degree of freedoms. We cannot exclude that this assumption leads to errors as AIMD simulation reveals numerous rotational movements of the adsorbed water. This effect influences the entropy change and stabilizes the non dissociated $A_1$ state with respect to the dissociated one $D_1 D_3$. This thermodynamic analysis may also be consistent with the previous kinetic observation. Indeed, the thermodynamic stabilization of the non dissociated reactant states with AMS induces that water dissociation rate constants are significantly smaller with AMS than with hTST. 

\subsection{Analysis of AMS reactive trajectories.} 

In addition to computing reaction rates, we show in this section how the AMS method allows to sample reactive trajectories. The overall AMS trajectories lengths are in the order of $200~\mathrm{ps}$. Qualitatively speaking, some chemically relevant trends can be identified. We identify there are two pathways for the rotation $A_4 \rightarrow A_1$. The first, and the less likely one, is similar to the path identified by the NEB static approach. The second one seems to be more similar to a $A_4 \rightarrow A_2A_3 \rightarrow A_1$ rotation, where the trajectory does not actually enter the $A_2A_3$ state but approaches it for a few femtoseconds before continuing toward the $A_1$ state. The same type of paths are observed in the few trajectories where a transition $D_1D_3 \rightarrow A_2 A_3$ occurs. However, such a systematic analysis of each reactive trajectory might become rapidly tedious and not safe enough to capture the overall chemical trends, since more than 2000 $A_1 \rightarrow D_1D_3$ trajectories are sampled by the AMS algorithm. An automated method is therefore necessary to analyze all of them and some dimensionality reduction is useful to this end. 

\begin{figure}[h!]
	\centering
	\includegraphics[width=12cm]{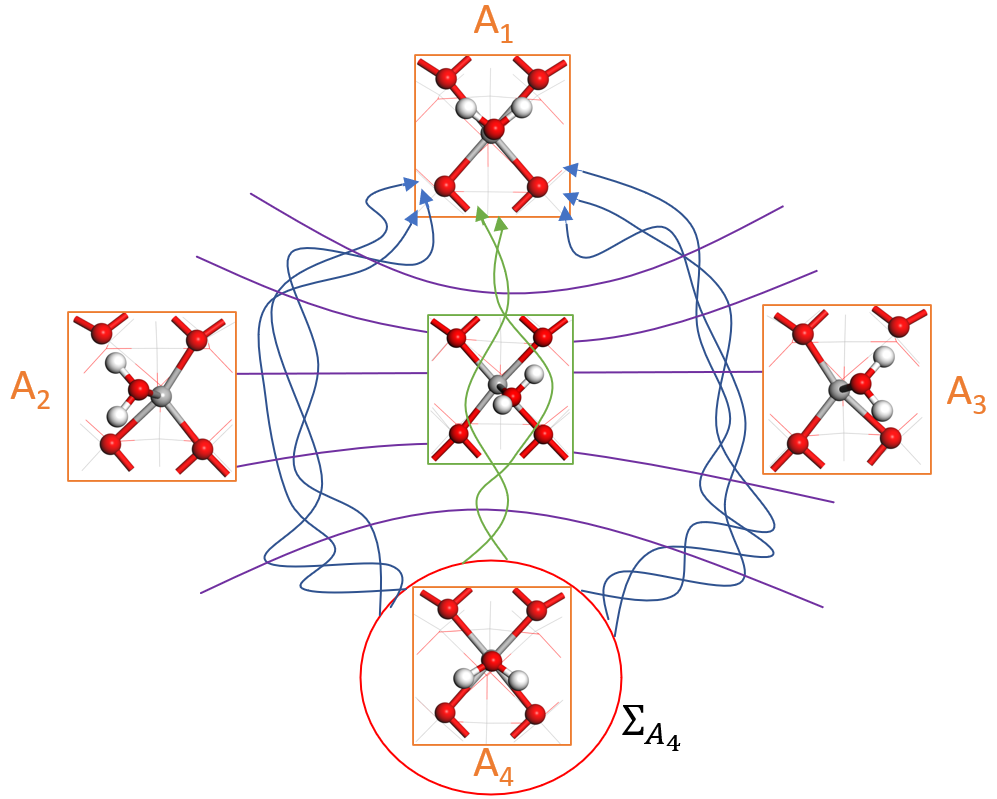}
\caption{Schematic representation of the two types of paths for the $A_4 \rightarrow A_1$ rotation. The first path (blue) is named "side" while the second one (green) is named "top". The purple line represent the RC isolevels used to represent the trajectories.}
	\label{reactraj_A4A1}
\end{figure}

\paragraph{Clustering reactive trajectories.} In the case of the $A_4 \rightarrow A_1$ rotation, two paths exist which can be identified by a visual inspection of many reactive trajectories. A more systematic way to proceed would be to rely on clustering methods, which are specially designed to identify groups within a dataset. Among the various possible approaches, we used here an approach based on the K-means algorithm as implemented in \textit{scikit-learn}\cite{Pedregosa2012}. To make numerical representation of each trajectory independant on its length, each trajectory was represented as the intersections of the trajectory and five isolevels of the $A_4$-vs-all SOAP-SVM RC. The details of the procedure to perform this clustering are presented in SI~Section~7. It is important to mention that K-means method requires to know \textit{a priori} the number of clusters to find thus various values should be tested. The two types of paths can be identified by visual inspection of the trajectory closest to each cluster's centroid even though all the trajectories are not perfectly assigned by this approach. Of course resorting to other clustering methods could be more efficient but such a systematic study is beyond the scope of the present work. The "top" path (see Figure~\ref{reactraj_A4A1} and trajectories supplied in electronic supplementary materials) qualitatively looks similar to the path found by the NEB. The fact that this path is less sampled than the "side" path indicates that this transition is rarer. 

\paragraph{Stochastic Transition State estimation.} One possibility is to consider only one structure per trajectory instead of the whole trajectory. The most important structure $\textbf{q}$ along a trajectory can be defined as the one such that the committor probability is $p_{R \rightarrow P}(\textbf{q}) = 0.5$, (where $p_{R \rightarrow P}(\textbf{q})$ is the probability that a molecular dynamics trajectory starting from $\textbf{q}$ reaches first the $P$ state rather than $R$. According to the IUPAC goldbook\cite{TS_goldbook2014}, in the part of the TS definition referring to a surface, all the structures satisfying the $p_{R \rightarrow P}(\textbf{q}) = 0.5$ conditions are part of the transition state. This definition of transition state as a "set of states (each characterized by its own geometry and energy)" is indeed not consistent with the following part of the definition "The transition state is characterized by one and only one imaginary frequency" which presents it as a first order saddle point on the potential energy surface. The various structures $\textbf{q}$ such that $p_{R \rightarrow P}(\textbf{q}) = 0.5$ are not necessarily identical to the saddle points identified via the NEB method and harmonic frequencies calculations, although some resemblance is expected. We propose to investigate this point in what follows for one water dissociation on the alumina surface.

As mentioned in the methods section, the estimated probability for a trajectory to reach the surface $\Sigma_{z_{\mathrm{kill}}^{n+1}}$ starting on the surface $\Sigma_{z_{\mathrm{kill}}^{n}}$ is $1 - \frac{\eta_{\mathrm{killed}}^{n+1}}{N_{\mathrm{rep}}}$. By identifying the level $n_{0.5}$ such that
\begin{equation}
\label{committor_05_estimator_AMS}
\prod_{n = n_{0.5}}^{n_\mathrm{max}} \left(1 - \frac{\eta_{\mathrm{killed}}^{n}}{N_{\mathrm{rep}}} \right)  = 0.5,
\end{equation}
one can define the iso-level $\Sigma_{0.5}$ of the reaction coordinate. The configurations corresponding to reactive trajectories crossing this surface are such that $\widehat{p}_{R \rightarrow P}(\textbf{q}) = 0.5$. There should be at least one structure corresponding to this condition per reactive trajectory. Considering only the first structure crossing the iso-level $\Sigma_{0.5}$, the mean structure is computed, in the sense of the SOAP descriptor. This analysis was applied for the various realizations of AMS that were run.

\begin{figure}
	\centering
	\includegraphics[width=12cm]{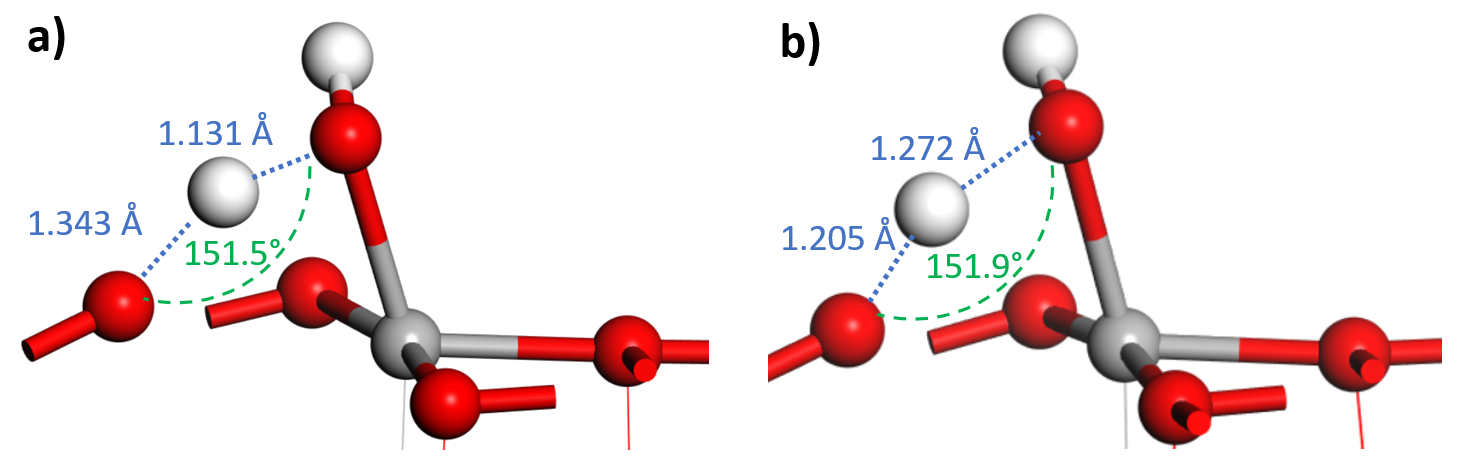}
\caption{Ball and sticks representation of \textbf{a)} saddle point on the PES and \textbf{b)} mean structures such that $p_{R \rightarrow P}(\textbf{q}) = 0.5$ on AMS using interpolated SOAP PCV RC. Color legend, red: oxygen, gray: aluminum, white: hydrogen.}
	\label{neb_vs_ams_ts}
\end{figure}

\paragraph{Stochastic Transition State of water dissociation.} For the dissociation event $A_1 \rightarrow D_1 D_3$, the interpolated SOAP-PCV reaction coordinate with the reactant and product states defined as $R~=~A_1~\cup~A_2A_3~\cup~A_4~\cup~D_2D_4$ and $P=D_1D_3$ conducts to a mean structure of configuration such that $p_{R \rightarrow P}(\textbf{q}) = 0.5$ qualitatively similar to the saddle point of the PES determined with the NEB method, as represented in Figure~\ref{neb_vs_ams_ts} (the corresponding trajectory is provided in electronic supplementary information). From a quantitative viewpoint, some slight structural differences can be noted regarding the O-H distances involving the transferred H atom. For the saddle point, the broken O-H bond is $0.14~\mathrm{\AA}$ shorter than for AMS, whereas the newly formed O-H bond is $0.14~\mathrm{\AA}$ larger. This difference might come from the fact that the momenta can bear a certain importance in the committor. Indeed the committor values estimated bears dynamical information while the saddle point is defined only with the positions. 

\begin{figure}[h!]
	\centering
	\includegraphics[width=10cm]{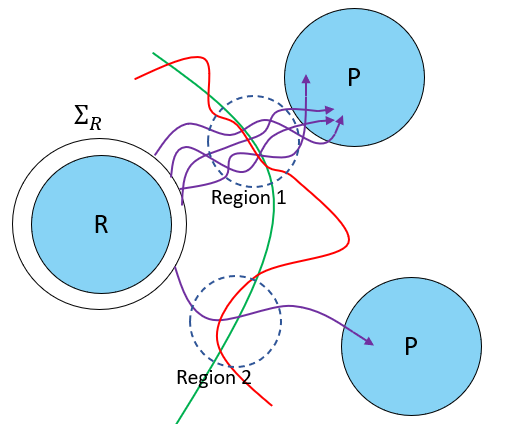}
\caption{Schematic representation of poor match of the $\Sigma_{0.5}$ iso-level of the committor function (red) and the reaction coordinate (green). The green iso-level is placed after an AMS sampling of some reactive trajectories from $R$ to $P$ where a majority of the trajectories has gone via Region 1.}
	\label{comittor_05_isolevel}
\end{figure}

The quality of this analysis depends on the quality of the sampling of the reaction path. Indeed considering the reactive trajectories sampled from the AMS done with: $R~=~A_1$ and $P=\cup~A_2A_3~\cup~A_4~\cup~D_1D_3~\cup~D_2D_4$ the definition of the stochastic TS is of poor quality. This comes from the fact that this AMS mostly samples $A_1 \rightarrow A_2A_3$ trajectories (and only rarely $A_1 \rightarrow D_1D_3$ trajectories). The best approximation of a stochastic TS is on the most sampled path (Region 1 in Fig.~\ref{comittor_05_isolevel}). This is in line with the obtained results for the confidence interval of the reaction rate constants (see Table~\ref{cv_table}). In Figure~\ref{comittor_05_isolevel}, the green curve represents the $\Sigma_{0.5}$ iso-level of the reaction coordinate, while the red one is the $\Sigma_{0.5}$ iso-level of the committor function. As these levels do not match perfectly on the whole space, the best approximation of the stochastic TS is in the region of space where most of the reactive trajectories concentrate (Region 1). This issue in the analysis of the reactive trajectories of less probable transitions is recurrent when multiple paths are sampled. An alternative approach to automatically identify whether multiple paths leading to a single product are present within a set of sampled trajectories would be desirable. 

\section{Conclusion}

We proposed and implemented a theoretical approach based on the Hill relation to compute the exact reaction rate constants using rare event sampling and support vector machines. It is illustrated on various chemical events occurring at an oxide material surface. A key algorithm to this end is the Adaptive Multilevel Splitting, which estimates reaction probabilities and samples reactive trajectories by using ab initio molecular dynamics. For that purpose, SVM was used to define the chemically relevant states and reaction coordinates to index the transition from reactant to products. It allows to compute the exact reaction rates for the dynamics at hand and makes possible a detailed analysis of reaction mechanisms via the inspection of reactive trajectories. The implementation done so as to communicate with a plane-wave DFT software allowed to illustrate the approach by studying the reactivity of a water molecule adsorbed on the $\gamma$-alumina $(100)$ surface. The computed reaction rate constants were discussed and compared to those of a static hTST approach. The methods precision is impacted by the choice of reaction coordinate, the choice of reactants and products in a multiple state situation, the number of repetitions of the probability estimation and the number of replicas intrinsic to AMS. The hTST approach does not make assumptions on the system's dynamics, but relies on strong assumptions concerning the shape of the potential energy surface, implying uncontrolled approximation of entropy. The proposed methodology allows to alleviate these limitations at the expense of an increased computational cost. 
Assuming that the Langevin dynamics accurately models the system's dynamics (which involves in particular having a relevant value of the friction coefficient), the presented approach should be more precise than TST approaches. In the case considered here, hTST reaction rate constants are always higher than the ones estimated via AMS and the Hill relation. The relative stability of states is also different. In particular, we show that hTST underestimates the thermodynamic stability of adsorbed water molecules, and simultaneously overestimates rate constants of water dissociation and formation. On top of that, the analysis of reactive trajectories allows to identify possible paths that are not clearly identified via the NEB approach.
Finally, this method used in combination with \textit{ab-initio} molecular dynamics can be computationally expensive. This issue might be alleviated by the use of Machine Learning Force Fields (MLFF), which can approach the accuracy of DFT force calculations at a much smaller computational cost. It also provides the opportunity to accurately describe nuclear quantum effects using path integral molecular dynamics such as in Ref.~\citenum{Bocus2023}. Some active learning schemes to train MLFFs have been proposed recently and they could articulate well with the present method\cite{Vandermause2020, Jinnouchi2020}. In particular, in contrast to standard MD, the presented approach favor a sampling of transition regions which are crucial to the description of chemical event. The study of a specific system could be done by first using jointly AMS and active learning to generate an accurate MLFF. Then, it could be used to evaluate accurately reaction rate constants and sample reactive trajectories. 

\begin{acknowledgement}

This project was realized in the framework of the joint laboratory IFPEN–Inria Convergence HPC/AI/HPDA for the energetic transition. Calculations were performed using the following HPC resources: Jean Zay and Occigen from GENCI-CINES, Joliot-Curie (Irene)  from TGCC/CEA (Grant A0120806134), ENER440 from IFP Energies nouvelles and Topaze from CCRT-CEA. The work of TL and GS was funded in part by the European Research Council (ERC) under the European Union's Horizon 2020 research and innovation programme (project EMC2, grant agreement No 810367).

\end{acknowledgement}


\bibliography{ams_alumina}

\end{document}

%\begin{tocentry}
%
%Some journals require a graphical entry for the Table of Contents.
%This should be laid out ``print ready'' so that the sizing of the
%Inside the \texttt{tocentry} environment, the font used is Helvetica
%8\,pt, as required by \emph{Journal of the American Chemical
%The surrounding frame is 9\,cm by 3.5\,cm, which is the maximum
%permitted for  \emph{Journal of the American Chemical Society}
%graphical table of content entries. The box will not resize if the
%content is too big: instead it will overflow the edge of the box.
%
%This box and the associated title will always be printed on a
%separate page at the end of the document.
%
%\end{tocentry}


\newpage 
\section{Multilevel splitting estimator and AMS pseudo code}
\label{SI_pseudo_code}

The various approaches using the Hill relation to compute reaction rate constants require a method to estimate the probability $p_{R \rightarrow P}(\partial R)$ to reach $P$ before $R$, starting from a given distribution on the boundary $\partial R$ of the reactant state $R$. This probability is often estimated using a splitting estimator such as FFS or AMS. Let us first explain why a naive Monte Carlo estimator is plagued by a large variance, before presenting the AMS estimator, and a pseudo code of the AMS algorithm. We also refer to the main text for a detailed explanation of the main steps of the algorithm. 

As observing a reaction is a rare event, the probability $p_{R \rightarrow P}(\partial R)$ is typically very small which is why resorting to a simple Monte-Carlo estimator is in general not efficient. A naive Monte Carlo estimator consists in running $n$ trajectories starting on the boundary $\partial R$ of $R$ and stopping them once they reach either the state $R$ or the state $P$. Counting the number $n_\mathrm{success}$ of trajectories which reach $P$ before $R$ ($R \rightarrow P$ transitions) yields the Monte-Carlo estimator: 
\begin{equation}
    \label{SI:monte_carlo_estimator}
    \widehat{p}_{R \rightarrow P}(\partial R) = \frac{n_{\mathrm{success}}}{n}.
\end{equation}
The normalised variance asscoiated with this estimator writes:
\begin{equation}
    \label{SI:monte_carlo_relative_error}
    \mathrm{Var}\left( \frac{\widehat{p}_{R \rightarrow P}(\partial R)}{p_{R \rightarrow P}(\partial R)}\right) = 
    \frac{(1-p_{R \rightarrow P}(\partial R))p_{R \rightarrow P}(\partial R)}{n (p_{R \rightarrow P}(\partial R))^2 } \approx \frac{1}{n p_{R \rightarrow P}(\partial R)},
\end{equation}
as $p_{R \rightarrow P}(\partial R)$ is negligible compared to $1$. From Equation~\eqref{SI:monte_carlo_relative_error}, it is clear that the lower the transition probability is, the larger the number of trials is needed to obtain a sensible relative error. 

To alleviate such a difficulty, a splitting estimator uses a product of conditional probabilities to reformulate the problem. The idea is to include the event of interest into an increasing sequence of more likely events. The target probability is then written as a product of conditional probabilities. More precisely, by introducing~$M$ surfaces $(\Sigma_j)_{1 \le j \le M}$ between~$R$ and~$P$ such that any transition path from $R$ to $P$ has to cross each of these surfaces, the probability~$p_{R \rightarrow P}^{\partial R}$ for the trajectory to reach $P$ before $R$, starting from the boundary of $R$ can be estimated as: 
\begin{equation}
\label{splitting_event}
\widehat{p}_{R \rightarrow P}(\partial R) =  \widehat{p}_{R \rightarrow \Sigma_1}(\partial R) \left( \prod_{j = 1}^{M-1} \widehat{p}_{R \rightarrow \Sigma_{j+1}}(\Sigma_{j}) \right) \widehat{p}_{R\rightarrow P}(\Sigma_{M})
\end{equation}
where $\widehat{p}_{R \rightarrow \Sigma_1}(\partial R)$ is an estimator of the probability for the path to reach $\Sigma_1$ before going back to~$R$, $\widehat{p}_{R \rightarrow \Sigma_{j+1}}(\Sigma_{j})$ is an estimator of the probability to reach $\Sigma_{j+1}$ before going back to $R$ conditionally on the fact that~$\Sigma_{j}$ was reached before going back to~$R$, and finally $\widehat{p}_{R\rightarrow P}(\Sigma_{M})$ is an estimator of the probability to reach $P$ before going back to $R$ conditionally on the fact that~$\Sigma_{M}$ was reached before going back to $R$.
It can be shown that such an estimator has a smaller variance than the Monte Carlo estimator. Moreover, for a fixed number of surfaces $M$, it can be shown that the variance is minimal if the surfaces are chosen such that all the conditional probabilities $p_{R \rightarrow \Sigma_{j+1}}(\Sigma_{j})$ are equal. This leads to the adaptive multilevel splitting (AMS) algorithm, where the surfaces are placed adaptively on a given simulation so that the estimator of these conditional probabilities are all equal, using empirical quantiles\cite{Brehier2015}. 

As an illustration, imagine than the probability to be estimated is $(1/2)^{M+1}$ (left-hand side of~\eqref{splitting_event}), and that the surfaces are positioned so that all the probabilities to be estimated in the product in the right-hand side are $1/2$ (there is 50\% chance to reach the next surface~$\Sigma_{j+1}$ before $R$, knowing that the path has reached $\Sigma_j$ before $R$). In such a situation, a naive Monte Carlo estimator is plagued by a large variance since the probability~$(1/2)^{M+1}$ to be estimated is very small. On the other hand, estimating $(M+1)$ times a probability of $1/2$ is much easier.

\begin{figure}[h!]
	\centering
	\includegraphics[width=12cm]{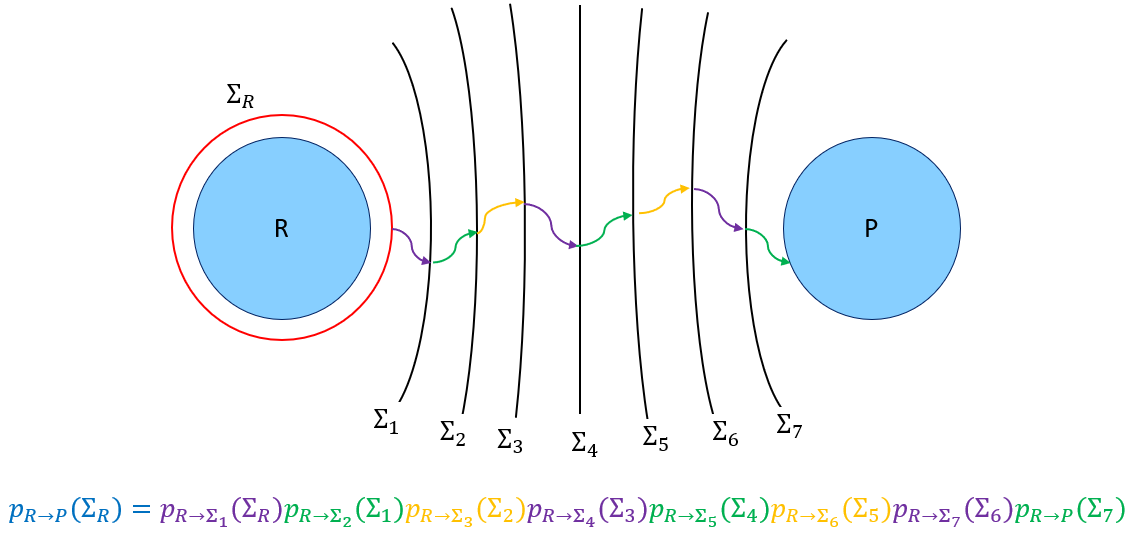}
\caption{Schematic representation of a splitting estimator.}
	\label{MLS_estimator}
\end{figure}

An illustration of this estimator with 7 surfaces is presented in Figure~\ref{MLS_estimator}. To use such an estimator in practice, one  needs to define the number of surfaces, their positions in phase space and to devise a way to estimate all the conditional probabilities $p_{R \rightarrow \Sigma_{j+1}}(\Sigma_{j})$. The AMS algorithm is designed to solve these problems all at once. 

The complete pseudo code of the AMS algorithm is  presented in Algorithm~\ref{SI:AMS_algo}, see the main text for the explanations of each steps.

\begin{algorithm}
	\textbf{Requires}\;
	$N_\mathrm{rep}$, $k_\mathrm{min}$, numerical definition of the states $R$ and $P$,  Reaction coordinate $\xi$,  $N_\mathrm{rep}$ initial conditions $\lbrace \textbf{q}_\mathrm{ini}^j \rbrace_{1 \leq j \leq N_{\mathrm{rep}}}$ on $\Sigma$ , Molecular dynamics engine $\textbf{MD\_step}$, \textbf{argsort} function returning the permutation of indices to sort an array of scalars, \textbf{randpick} functions that randomly picks an element of an array\;
	\KwOut{
		\\ $p$: the estimated probability of reaching $P$ before $R$ starting from $\Sigma$.
	}
	\textbf{Algorithm}\;
	$p \leftarrow 1$\;
	\For{$i=1$ \normalfont{\textbf{to}} $i=N_\mathrm{rep}$}{
		$\textbf{q}_0^{i} \leftarrow \textbf{q}_\mathrm{ini}^i$\;
		$t \leftarrow 0$\;
		\While{${\textbf{q}}_t^{i}\not\in R \cup P$}{
			$\textbf{q}_{t + \delta t}^{i} \leftarrow \textbf{MD\_step}(\textbf{q}_t^{i})$\;
			$t \leftarrow t + \delta t$\;
		}
		$z_{\mathrm{max}}^{i} \leftarrow \underset{t}{\sup}(\xi (\textbf{q}_t^{i}))$\;
	}
	\While{$\exists \; i \in \llbracket 1, N_\mathrm{rep} \rrbracket \; \big/ \; z_\mathrm{max}^i \neq +\infty$}{
		sorted $\leftarrow$  \textbf{argsort}$\left(z_{\mathrm{max}}\right)$\;
		$z_\mathrm{kill} \leftarrow  z_{\mathrm{max}}^{\mathrm{sorted[k_\mathrm{min}]}}$\;
		killed $\leftarrow \left\{ i  \in \llbracket 1, N_\mathrm{rep} \rrbracket \; \middle| \; z_{\mathrm{max}}^{i} \leq z_\mathrm{kill} \right\}$\;
		alive $\leftarrow \left\{ i  \in \llbracket 1, N_\mathrm{rep} \rrbracket \; \middle| \; z_{\mathrm{max}}^{i} > z_\mathrm{kill} \right\}$\;
		$p \leftarrow p\left(1 - \frac{\mathrm{\textbf{length}(\mathrm{killed})}}{N_\mathrm{rep}} \right)$\;
		\For{$k \in \mathrm{killed}$ }{
			\textbf{delete} $\left(\textbf{q}_{t}^{k}\right) \; \forall t$\;
			j $\leftarrow \mathrm{\textbf{randpick}}(\mathrm{alive})$\;
			$t \leftarrow 0$\;
			\While{$\xi\left(\normalfont{\textbf{q}}_{t}^{k}\right) \leq z_\mathrm{kill}$}{
				$\textbf{q}_{t + \delta t}^{k} \leftarrow \textbf{q}_{t + \delta t}^{j}$\;
				$ t \leftarrow t + \delta t$\;
			}
			\While{$ \normalfont{\textbf{q}}_t^{k} \not \in R \cup P$}{
				$\textbf{q}_{t + \delta t}^{k} \leftarrow \textbf{MD\_step}(\normalfont{\textbf{q}}_t^{k})$ \;
				$t \leftarrow t + \delta t$\;
			}
			$z_{\mathrm{max}}^{k} \leftarrow \underset{t}{\sup}(\xi (\textbf{q}_t^{k}))$\;
		}
	}
	\Return{$p$}
	\caption{\label{SI:AMS_algo}
		Simplified AMS pseudo algorithm}
\end{algorithm}

\newpage 
\section{Rate constant error estimation}
\label{SI_error_estimation}

We provide in this section, an expression for the confidence interval for the reaction rate using the Delta method, which is a standard technique in statistics\cite{Hogg}.
The Hill relation writes: 
\begin{equation}
\label{SI:Hill_relation}
k_\mathrm{Hill} = \Phi_R p_{R \rightarrow P}(\partial R),
\end{equation}
where~$\Phi_R$ is the flux of trajectories leaving the state~$R$ and $p_{R \rightarrow P}$ is the probability of reaching~$P$ before~$R$ starting on the boundary of $R$. The flux in~\eqref{SI:Hill_relation} is estimated via: 
\begin{equation}
\label{SI:flux_approx}
\Phi_R = \frac{n_{\mathrm{loop}-R\Sigma_R R }}{t_{\mathrm{tot}}} = \frac{1}{t_{\mathrm{loop}-R\Sigma_R R }}.
\end{equation}
To obtain uncorrelated loop times $t_{\mathrm{loop}-R\Sigma_R R }$, the results presented were computed considering one loop every five loops. From~(\ref{SI:Hill_relation}) and~(\ref{SI:flux_approx}), the reaction rate writes: 
\begin{equation}
\label{SI:approx_transition_rate2}
k_{R \rightarrow P} = \frac{p_{R \rightarrow P}(\Sigma_R)}{t_{\mathrm{loop}-R\Sigma_R R}}.
\end{equation}
Assuming $M_\mathrm{real}$ realizations of AMS were done, let us consider the two estimators of the term of the quotient: 
\begin{equation}
\label{SI:mean_proba_time}
\begin{aligned}
\widehat{p}_{R \rightarrow P}(\Sigma_R) &= \frac{1}{M_\mathrm{real}} \sum_{i=1}^{M_\mathrm{real}} p_i,\\
\widehat{t}_{loop-R\Sigma_R R} &= \frac{1}{n_{\mathrm{loop}-R\Sigma_R R}} \sum_{j=1}^{n_{\mathrm{loop}-R\Sigma_R R}} t_i,
\end{aligned}
\end{equation}
where $p_i$ are independent results of AMS with the same $N_\mathrm{rep}$ and $k_\mathrm{min}$, the times $t_i$ are the times of different loops going from $R$ to $\Sigma$ and then back to $R$.

The AMS estimator satisfies the central limit theorem in the limit of an infinitely large number of replicas, see Ref. \citenum{Cerou2019}. Concerning the estimator of this flux, the central limit theorem can be invoked only if the times $t_i$ are not correlated. It is clear that two successive times might be correlated but due to the fact that the Langevin dynamics is stochastic, it is possible to assume that $t_i$ and $t_{i+n}$ are not correlated if~$n$ is large enough. The number of times $n$ one should skip to ensure that depends on the friction parameter of the dynamics and was assumed to be~$5$ in this work. Assuming that the central limit theorem holds, one obtains
\begin{equation}
\begin{aligned}
\widehat{t}_{\mathrm{loop}-R\Sigma_R R} & = t_{\mathrm{loop}-R\Sigma_R R} +\sqrt{\frac{\mathrm{Var}(\widehat{t}_{\mathrm{loop}-R\Sigma_R R})}{n_{\mathrm{loop}-R\Sigma_R R}}}G_t,\\
\widehat{p}_{R \rightarrow P}(\Sigma_R)& = p_{R \rightarrow P}(\Sigma_R) + \sqrt{\frac{\mathrm{Var}(\widehat{p}_{R \rightarrow P}(\Sigma_R))}{M_\mathrm{real}}} G_p
\end{aligned}
\end{equation}
where $G_t$ and $G_p$ are two real valued random variables distributed according to a standard Gaussian distribution. By truncation of the Taylor expansion at the first order in $\frac{1}{\sqrt{M_\mathrm{real}}}$ and $\frac{1}{\sqrt{n_{\mathrm{loop}-R\Sigma_R R}}}$, one gets: 
\[
\widehat{k}_{R \rightarrow P}  \approx \frac{p_{R \rightarrow P}(\Sigma_R)}{t_{\mathrm{loop}-R\Sigma_R R}} + \frac{1}{t_{\mathrm{loop}-R\Sigma_R R}}\sqrt{\frac{\mathrm{Var}(\widehat{p}_{R \rightarrow P}(\Sigma_R))}{M_\mathrm{real}}}G_p  - \frac{p_{R \rightarrow P}(\Sigma_R)}{t_{\mathrm{loop}-R\Sigma_R R}^2}\sqrt{\frac{\mathrm{Var}(\widehat{t}_{\mathrm{loop}-R\Sigma_R R})}{n_{\mathrm{loop}-R\Sigma_R R}}} G_t.
\]
As the sum of two zero mean Gaussian random variables is also a zero mean Gaussian random variable, it therefore holds: 
\begin{equation}
\label{SI:rate_estimator}
\widehat{k}_{R \rightarrow P}  \approx k_{R \rightarrow P}  + \sqrt{\frac{\mathrm{Var}(\widehat{p}_{R \rightarrow P}(\Sigma_R))}{t_{\mathrm{loop}-R\Sigma_R R}^2 M_\mathrm{real}}  + \frac{p_{R \rightarrow P}(\Sigma_R)^2\mathrm{Var}(\widehat{t}_{\mathrm{loop}-R\Sigma_R R}) }{t_{\mathrm{loop}-R\Sigma_R R}^4 n_{\mathrm{loop}-R\Sigma_R R}}}G_k.
\end{equation}

Using the unbiased variance estimators
\begin{equation}
\label{SI:var_proba_time}
\begin{aligned}
\mathrm{Var}(\widehat{p}_{R \rightarrow P}(\Sigma_R)) & =  \frac{1}{M_\mathrm{real} - 1} \sum_{i=1}^{M_\mathrm{real}} \left(p_i - \widehat{p}_{R \rightarrow P}(\Sigma_R)\right)^2, \\ 
\mathrm{Var}(\widehat{t}_{\mathrm{loop}-R\Sigma_R R}) & = \frac{1}{n_{\mathrm{loop}-R\Sigma_R R} - 1} \sum_{i=1}^{n_{\mathrm{loop}}} \left(t_i - \widehat{t}_{loop-R\Sigma_R R}\right)^2,
\end{aligned}
\end{equation}
and replacing $t_{loop-R\Sigma_R R}$ and ${p}_{\Sigma_R \rightarrow P}$ by their estimators in~(\ref{SI:rate_estimator}), the following confidence interval are finally deduced:
\begin{equation}
\label{SI:final_conf_interval}
\begin{aligned}
k_{R \rightarrow P} & \in \left[\frac{\widehat{p}_{R \rightarrow P}(\Sigma_R)}{\widehat{t}_{\mathrm{loop}-R\Sigma_R R}} - \theta_\frac{\alpha}{2} \sigma, \frac{\widehat{p}_{R \rightarrow P}(\Sigma_R)}{\widehat{t}_{\mathrm{loop}-R\Sigma_R R}} + \theta_\frac{\alpha}{2}\sigma \right],\\
\sigma & = \sqrt{\frac{\mathrm{Var}(\widehat{p}_{R \rightarrow P}(\Sigma_R))}{\widehat{t}_{\mathrm{loop}-R\Sigma_R R}^2 M_\mathrm{real}}  + \frac{\widehat{p}_{R \rightarrow P}(\Sigma_R)^2\mathrm{Var}(\widehat{t}_{\mathrm{loop}-R\Sigma_R R}) }{\widehat{t}_{loop-R\Sigma_R R}^4 n_{\mathrm{loop}-R\Sigma_R R}}},
\end{aligned}
\end{equation}
where $\theta_\frac{\alpha}{2}$ stand for the quantile $\frac{\alpha}{2}$ of the Gaussian law to obtain a $1 -alpha$ precision confidence interval.
\newpage 
\section{State to state probability estimation in a multi-state case}
\label{SI_multistate}

Only two states $R$ and $P$ are necessary for AMS while multiple states are generally present in catalysis and the reaction rate constants between all the states $\{E_1, .. \,E_i, .. \, E_N\}$  are of interest. To address this issue with AMS, two approaches can be proposed. 

\paragraph{First approach.} We first define $R = E_j$ and $P = \underset{i \neq j}{\bigcup} E_i$.  The initial conditions are sampled on the surface $\Sigma_{E_j}$ surrounding the state $E_j$, then the transition probability $\Sigma_{E_j} \rightarrow P$ can be estimated. Finally the probabilities $\Sigma_{E_j} \rightarrow E_i$ can be estimated by counting the number of trajectories ${n_{E_i}^\mathrm{in}}$ that indeed finished in the $E_i$ state. 
\begin{equation}
\label{SI:states_probas}
\widehat{p}_{E_j \rightarrow E_i}(\Sigma_{E_j}) = \frac{n_{E_i}^\mathrm{in}}{N_\mathrm{rep}} \widehat{p}_{E_j \rightarrow P}(\Sigma_{E_j}).
\end{equation}
This formula is motivated later on by considering the order of the first hitting time of each state. This approach allows to observe various types of transitions using a single AMS run. However if one transition is less likely to occur compared to another, the sampling of the less probable transition might not be satisfactory as most trajectories would sample the most probable one. 

\paragraph{Second approach.} To circumvent this issue, one can decide to change the definition of the reactant and product states. Using initial conditions sampled on $\Sigma_{E_j}$ and setting $R = \underset{i \neq k}{\bigcup} E_i$ and $P = E_k$, the AMS is compelled to sample the $E_j \rightarrow E_k$ trajectories. The state $R$ contains states $E_i$ with $i \neq k$ as it allows to consider a $E_j \rightarrow E_i$ as a non reactive trajectory. This matter is discussed in the result section. 

\paragraph{Justification of equation~\eqref{SI:states_probas}.} Let us consider a 3 states case. The reactant state is $R = E_1$ and the product state is $P = E_2 \cup E_3$. Then we define the time $\tau_{E_i}$ as the first that the dynamics starting on $\Sigma_R$ reaches the state $E_i$. As three states are considered, six possibilities for the ranking of these three time are possible: 
\begin{enumerate}
	\item $\tau_{E_1} < \tau_{E_2} < \tau_{E_3}$,
	\item $\tau_{E_1} < \tau_{E_3} < \tau_{E_2}$,	
	\item $\tau_{E_2} < \tau_{E_1} < \tau_{E_3}$, 
	\item $\tau_{E_2} < \tau_{E_3} < \tau_{E_1}$, 
	\item $\tau_{E_3} < \tau_{E_1} < \tau_{E_2}$, 
	\item $\tau_{E_3} < \tau_{E_2} < \tau_{E_1}$.
\end{enumerate} 
Hence, the sum of the probability of these 6 events in equals to 1 and, given that the states do not overlap, these events are independent. It is possible to identify that the event corresponding to reaching $P$ before $R$ correspond the events 3 to 6 while the events 1 and 2 correspond to reaching $R$ first. Reaching $E_2$ before $R$ corresponds to the events 3 and 4 and finally reaching $E_3$ before $R$ corresponds to events 5 and 6. Then, to estimate the probability of the last two events one has just to identify the fraction of events 3 and 4 (or 5 and 6) that occurred during the realization of the events 3 to 6. This is exactly what the factor $ \frac{n_{E_i}^\mathrm{in}}{N_\mathrm{rep}}$ represents in expression~(\ref{SI:states_probas}).
\newpage 
\section{Calculation parameters}
\label{SI_calculation_parameters}

\subsection{DFT parameters}
\label{SI_DFT_parameters}

The DFT functional was PBE\cite{Perdew1996} with the D3 dispersion correction\cite{Grimme2010}. A Gaussian smearing was used with $\sigma= 0.05 \; \mathrm{eV}$. The $\gamma$-alumina bulk structure taken from Ref. \citenum{Krokidis2001}. The K point grid was set to $2\times2\times4$ and centered at the $\Gamma$ point. The bulk structure was first fully relaxed (allowing box volume to change) with a 800 eV kinetic energy cutoff to ensure low Pulay stress. All other DFT calculations on slabs representing the $\gamma$-alumina (100) surface were achieved by keeping the cell's volume constant with a kinetic energy cutoff of 450 eV. The $(100)$ surface model is composed of a four layer slab structure with 15 $\mathrm{\AA}$ of vacuum inserted in the direction perpendicular to the surface plane (that is, the $x$ direction in this case). Calculations on this system were carried out with a K point grid set to $1\times2\times4$. Geometry optimizations were done using the conjugate gradient algorithm as implemented in VASP with a convergence criterion of $0.01 \;  \mathrm{eV/\mathrm{\AA}}$. 

\subsection{AIMD parameters.}
\label{SI_AIMD_parameters}
All the molecular dynamics runs were generated using the Brünger Brooks Karplus integrator of the Langevin dynamics implemented in VASP. A $1.0~\mathrm{fs}$ time step was used for all of them. The length of the dynamics runs and the used friction parameter varied upon the purpose of the molecular dynamics run as detailed in the results section of the main text. 

\subsection{NEB and saddle points}
\label{SI_NEB_parameters}
Saddle points on the potential energy surface were identified by nudged elastic band (NEB) methods using the VASP TST tools\cite{Jonsson1998, VTSTTools}. The spring force was taken to $5.0 \; \mathrm{eV.\AA}^{-2}$ and nudging was turned on. The number of images was 10 including the reactant and product. The optimizer used was FIRE with the default parameters that can be found in the VTST documentation \cite{VTSTTools}. The initial path was created by an interpolation between the z-matrix representation of the reactant and product structures with the Opt'n path code\cite{Optnpath}.  The identification of the relevant saddle points on the potential energy surface was done starting from the NEB results and refined using the quasi-Newton method. The vibrational frequencies of the minima and the saddle points were evaluated using a finite difference method as implemented in the VASP package based on displacements of $0.01~\mathrm{\AA}$. Using these frequencies, the free energies of the meta-stable basins and the transition states were computed within the harmonic approximation. The rotational components of the entropy were not explicitly computed and assumed to cancel out between minima and transition states. Detailed expressions used can be found in Ref. \citenum{Mcdouall2013}. Finally the hTST rates were computed using Eyring-Polanyi equation assuming that the transmission coefficient is equal to 1. 

\subsection{SOAP descriptor parameters}
\label{SI_SOAP_parameters}
The SOAP descriptors were computed using the dscribe python package\cite{Himanen2020}. The cutoff radius was set to 6 $\mathrm{\AA}$ as the main structural changes in the example system are within a sphere of this radius around a central atoms. The atomic density in the neighborhood of an atom is approximated as a sum of Gaussians centered on the nearby atomic nuclei and their width $\sigma$  was $0.05 \; \mathrm{\AA}$. The width of these Gaussians is chosen so that there is not too much overlap between two different structures. The parameters $n_\mathrm{max}$ and $l_\mathrm{max}$ controlling the size of the basis on which the atomic density is projected were respectively set to 8 and 6.  
\newpage 
\section{Implementation with VASP software}
\label{SI_implementation}

The Fleming-Viot particle process to sample initial conditions and the AMS were implemented in python scripts calling the execution of the VASP software for the integration of the unbiased Langevin dynamics. Both these algorithms require to stop the dynamics when it enters a certain state. This means that at every time-step, one has to evaluate the criterion used to define this state and then decide whether the dynamics is to be continued or not. This kind of stopping condition cannot be enforced with the current implementation of VASP and had to be implemented. The collective variable to define the states is computed using a python script \code{CV.py} that is used as input. The choice of the stopping conditions is monitored by INCAR tags. 

\newpage 
\section{Detailed numerical results}
\label{SI_detailed_num_results}

The most precise results for each observed transitions during this study are presented in Table~\ref{SI:all_rates_table}. From these results, some reaction heats were computed and presented in Table~\ref{SI:reaction_heat_tables}. 
\begin{table}
	\caption{Transition rate constants computed with AMS for all the transition observed in this study with 90\% precision.}
	\label{SI:all_rates_table}
	\begin{tabular}{llll}
	\hline
	Transition & $t_{\mathrm{loop}-R\Sigma_{R}R}$ (fs) & $p_\mathrm{Transition}$ & $k_\mathrm{Transition} (\mathrm{s}^{-1})$   \\
	\hline
        \multicolumn{4}{l}{Water rotations}\\
        \hline
        $A_1 \rightarrow A_2A_3$ \quad \textsuperscript{\emph{a}}   & $110 \pm 5$ & $\left(3.38\pm 1.56\right) \, 10^{-3}$ & $\left(3.08 \pm 1.43\right) \, 10^{10} $ \\
        $A_2A_3 \rightarrow A_1$\quad \textsuperscript{\emph{b}}   & $85 \pm 9$ & $\left(1.34\pm 0.41\right) \, 10^{-2}$ & $\left(1.49 \pm 0.46\right) \, 10^{11} $ \\
        \hline
        $A_2A_3 \rightarrow A_4$ \quad\textsuperscript{\emph{b}}   & $85 \pm 9$ & $\left(3.90\pm 1.97\right) \, 10^{-3}$ & $\left(4.33 \pm 2.20\right) \, 10^{10} $ \\
        $A_4 \rightarrow A_2A_3$ \quad \textsuperscript{\emph{c}}   & $100 \pm 5$ & $\left(2.24\pm 0.83\right) \, 10^{-2}$ & $\left(2.35 \pm 0.87\right) \, 10^{11} $ \\
        \hline
        $A_1 \rightarrow A_4$ \quad \textsuperscript{\emph{a}}      & $110 \pm 3$ & $\left(3.66\pm 7.18\right) \, 10^{-7}$ & $\left(3.34 \pm 6.56\right) \, 10^{6} $ \\
        $A_4 \rightarrow A_1$ \quad \textsuperscript{\emph{c}}      & $100 \pm 5$ & $\left(1.28\pm 0.65\right) \, 10^{-3}$ & $\left(1.34 \pm 0.68\right) \, 10^{10} $ \\
        \hline
        \multicolumn{4}{l}{Hydroxyl rotation}\\
        \hline
        $D_1D_3 \rightarrow D_2D_4$ \quad \textsuperscript{\emph{d}}& $\varnothing$ & $\varnothing$ & $\varnothing$ \\
        $D_2D_4 \rightarrow D_1D_3$ \quad \textsuperscript{\emph{e}}& $63 \pm 4$ & $\left(1.72\pm 2.84\right) \, 10^{-5}$ & $\left(2.86 \pm 4.71\right) \, 10^{8} $ \\
	\hline
        \multicolumn{4}{l}{Formation and dissociation of water} \\
        \hline
        $A_1 \rightarrow D_1D_3$ \quad \textsuperscript{\emph{f}}   & $109 \pm 3$ & $\left(1.78 \pm 1.64\right) \, 10^{-4}$ &  $\left(1.64 \pm 1.59\right) \, 10^{9} $ \\
        $D_1D_3 \rightarrow A_1$ \quad \textsuperscript{\emph{d}}   & $88 \pm 4$ & $\left(2.05\pm 1.40\right) \, 10^{-3}$ & $\left(2.32 \pm 1.59\right) \, 10^{10} $ \\
        \hline
        $A_2A_3 \rightarrow D_2D_4$ \quad \textsuperscript{\emph{b}}& $85 \pm 9$ & $\left(7.07\pm 6.77\right) \, 10^{-4}$ & $\left(7.86 \pm 7.53\right) \, 10^{9} $ \\
        $D_2D_4 \rightarrow A_2A_3$ \quad \textsuperscript{\emph{e}}& $63 \pm 4$ & $\left(7.71\pm 3.24\right) \, 10^{-3}$ & $\left(1.28 \pm 0.54\right) \, 10^{11} $ \\
        \hline
        $A_2A_3 \rightarrow D_1D_3$ \quad \textsuperscript{\emph{b}}& $\varnothing$ & $\varnothing$ & $\varnothing$ \\        
        $D_1D_3 \rightarrow A_2A_3$ \quad \textsuperscript{\emph{d}}& $88 \pm 4$ & $\left(2.05\pm 2.77\right) \, 10^{-5}$ & $\left(2.33 \pm 3.14\right) \, 10^{8} $ \\
	\hline
    \end{tabular}\\
\textsuperscript{\emph{a}} Rate sampled using $N_\mathrm{rep} = 200$, $M_\mathrm{real} = 10$, $R~=~A_1$, $P=~A_2A_3~\cup~A_4~\cup~D_1D_3~\cup~D_2D_4$ and $\xi  =$ $A_1$-vs-all SOAP-SVM RC. \\
\textsuperscript{\emph{b}} Rate sampled using $N_\mathrm{rep} = 100$, $M_\mathrm{real} = 20$, $R~=~A_2A_3$, $P=~A_1~\cup~A_4~\cup~D_1D_3~\cup~D_2D_4$ and $\xi  =$ $A_2A_3$-vs-all SOAP-SVM RC. \\
\textsuperscript{\emph{c}} Rate sampled using $N_\mathrm{rep} = 200$, $M_\mathrm{real} = 10$, $R~=~A_4$, $P=~A_1~\cup~A_2A_3~\cup~D_1D_3~\cup~D_2D_4$ and $\xi  =$ $A_4$-vs-all SOAP-SVM RC. \\
\textsuperscript{\emph{d}} Rate sampled using $N_\mathrm{rep} = 200$, $M_\mathrm{real} = 10$, $R~=~D_1D_3$, $P=~A_1~\cup~A_2A_3~\cup~A_4~\cup~D_2D_4$ and $\xi  =$ $D_1D_3$-vs-all SOAP-SVM RC. \\
\textsuperscript{\emph{e}} Rate sampled using $N_\mathrm{rep} = 200$, $M_\mathrm{real} = 10$, $R~=~D_2D_4$, $P=~A_1~\cup~A_2A_3~\cup~A_4~\cup~D_1D_3$ and $\xi  =$ $D_2D_4$-vs-all SOAP-SVM RC. \\
\textsuperscript{\emph{f}} Rate sampled using $N_\mathrm{rep} = 200$, $M_\mathrm{real} = 10$, $R~=~A_1~\cup~A_2A_3~\cup~A_4~\cup~D_2D_4$, $P=D_1D_3$, $\Sigma_R = \Sigma_{A_1}$ and $\xi  =$ interpolated SOAP-PCV
\end{table}

\begin{table}
	\caption{Reaction heats at 200 K computed AMS reaction rates of Table~\ref{SI:all_rates_table}} 
	\label{SI:reaction_heat_tables}
	\begin{tabular}{ll}
		\hline
		& Value $(\mathrm{kJ.mol}^{-1})$ \\
		\hline
        \multicolumn{2}{l}{Water rotations} \\
        \hline
		$\Delta G_{A_1 \rightarrow A_2 A_3} $     & $2.62 \pm 2.66$ \\
		$\Delta G_{A_2 A_4 \rightarrow A_4} $     & $2.81 \pm 2.83$ \\
        $\Delta G_{A_1 \rightarrow A_4} $         & $13.8 \pm 4.45$ \\
        \hline
        \multicolumn{2}{l}{Water dissociations} \\
        \hline
		$\Delta G_{A_2 A_3 \rightarrow D_2 D_4} $ & $4.64 \pm 3.54$ \\
        $\Delta G_{A_1 \rightarrow D_1 D_3} $     & $4.41 \pm 3.88$ \\
		\hline
	\end{tabular}
\end{table}

Similar results can be obtained via the hTST approach and are presented in Tables~\ref{SI:all_rates_table_hTST}~and~\ref{SI:reaction_heat_tables_h}. 

\begin{table}
	\caption{Activation energies and rate constant computed with the harmonic approximation at 200K.}
	\label{SI:all_rates_table_hTST}
	\begin{tabular}{lcc}
		\hline
		Transition & $\Delta G_\mathrm{Transition}^\ddagger \; (\mathrm{kJ.mol}^{-1}) $ & $k_\mathrm{Transition} (\mathrm{s}^{-1})$   \\
		\hline
        \multicolumn{3}{l}{Water rotations}\\
        \hline
	  $A_1 \rightarrow A_2A_3$   & $6.67$ & $7.55 \; 10^{10}$ \\
        $A_2A_3 \rightarrow A_1$   & $1.17$ & $2.06 \; 10^{12}$ \\
        \hline
        $A_2A_3 \rightarrow A_4$   & $7.88$ & $3.64 \; 10^{10}$ \\
        $A_4 \rightarrow A_2A_3$   & $3.31$ & $5.66 \; 10^{11}$ \\
        \hline
        $A_1 \rightarrow A_4$      & $16.5$ & $2.04 \; 10^{8}$ \\
        $A_4 \rightarrow A_1$      & $6.44$ & $8.65 \; 10^{10}$ \\
        \hline
        \multicolumn{3}{l}{Hydroxyl rotation}\\
        \hline
        $D_1D_3 \rightarrow D_2D_4$& $12.4$ & $2.38 \; 10^{9}$ \\
        $D_2D_4 \rightarrow D_1D_3$& $11.5$ & $4.15 \; 10^{9}$ \\
	\hline
        \multicolumn{3}{l}{Formation and dissociation of water} \\
        \hline
        $A_1 \rightarrow D_1D_3$   & $4.18$ & $3.37 \; 10^{11}$\\
        $D_1D_3 \rightarrow A_1$   & $2.17$ & $1.13 \; 10^{12}$\\
        \hline
        $A_2A_3 \rightarrow D_2D_4$& $-4.28$ & $5.45 \; 10^{13}$\\
        $D_2D_4 \rightarrow A_2A_3$& $-1.71$ & $1.17 \; 10^{13}$\\
        \hline
	\end{tabular}
	
\end{table}

\begin{table}
	\caption{Reaction heats computed from harmonic approximation of the free energy at 200 K } 
	\label{SI:reaction_heat_tables_h}
	\begin{tabular}{ll}
		\hline
		& Value $(\mathrm{kJ.mol}^{-1})$ \\
		\hline
        \multicolumn{2}{l}{Water rotations} \\
        \hline
		$\Delta G_{A_1 \rightarrow A_2 A_3} $     & $5.50$ \\
		$\Delta G_{A_2 A_4 \rightarrow A_4} $     & $4.56$ \\
        $\Delta G_{A_1 \rightarrow A_4} $         & $10,1$ \\
        \hline
        \multicolumn{2}{l}{Water dissociations} \\
        \hline
        $\Delta G_{D_1D_3 \rightarrow D_2D_4} $   & $0,93$ \\
        \hline
        \multicolumn{2}{l}{Water dissociations} \\
        \hline
		$\Delta G_{A_2 A_3 \rightarrow D_2 D_4} $ & $-2.56$ \\
        $\Delta G_{A_1 \rightarrow D_1 D_3} $     & $2.01$ \\
		\hline
	\end{tabular}
	
\end{table}

\newpage 
\section{Clustering reactive trajectories}
\label{SI_clustering_reactive_trajectories}

K-means and most clustering algorithms better perform for clustering problems in low dimensions. Performing clustering using the whole reactive trajectories is therefore expected to be inefficient. Moreover, the sampled reactive trajectories do not necessary have the same length and cannot directly be compared. To alleviate these problems, a first preprocessing step is to summarize each trajectory by a small number of structures. These structures correspond to the first point of the trajectory crossing certain reaction coordinate isolevels. In the example presented in the result section of the main text, five levels were used. These levels are equally spaced between the largest minimal values of the RC along the trajectories and the smallest maximal values of the RC along the trajectories. The next preprocessing step is to numerically represent these few structures per trajectory via the SOAP descriptor centered on the oxygen atom of the water molecule. As these SOAP descriptors are also in high dimension, a principal component analysis is performed for all the normalized SOAP descriptors of structures corresponding to the same level of the RC. Finally, the first four principal components are used as descriptors of the structure. The choice of using the first four principal components is motivated by the fact that these four components capture at least 90\% of the variance of the SOAP descriptors for a given level. Finally, a full reactive trajectory of is represented as a vector of size four times the number of levels chosen.  As the K-means algorithm strongly depends on its (random) initialisation, the algorithm was repeated 20 times and the best set of clusters was kept. Setting the number of clusters to find to 3 allows to find the two types of pathways for the $A_4 \rightarrow A_1$ rotation which were previously discussed. The trajectories are attached as videos in the electronic SI. Three clusters are necessary as both paths are not equally sampled by the AMS simulation. Indeed, one path being less probable, is less sampled. This behavior is typical of K-means which prefers to split one large cluster in two parts rather than identifying a large one and a much smaller one. Of course such issues could be alleviated by resorting to other clustering methods but such study is beyond the scope of the present work. However, the result obtained at this level paves the way to future more detailed investigations.

\newpage 



\bibliography{ams_alumina}